\documentclass[twocolumn,nofootinbib,10pt,tightenlines,superscriptaddress]{revtex4-2}

\usepackage{geometry}
\usepackage{microtype}

\renewcommand{\paragraph}[1]{\vspace{7pt}\noindent\textbf{#1}}

\usepackage{amsmath,amssymb}
\usepackage[hidelinks]{hyperref}
\usepackage{graphicx}
\usepackage{array}

\usepackage{enumitem}
\setlist{noitemsep}

\usepackage{natbib}
\setcitestyle{super}

\usepackage[label font=bf,labelformat=simple,caption=false]{subfig}
\usepackage{floatrow}
\renewcommand{\vec}[1]{\boldsymbol{\mathrm{#1}}}
\newcommand{\Diag}{\mathsf{Diag}}
\renewcommand{\P}{\mathcal{P}}
\newcommand{\C}{\mathcal{C}}

\usepackage{color}
\def\edcolor{black}
\newcommand{\ed}[1]{\textcolor{\edcolor}{#1}}
\newcommand{\edStart}{\color{\edcolor}}
\newcommand{\edEnd}{\color{black}}

\begin{document}

\title{Bridging the short-term and long-term dynamics of economic structural change}

\author{James McNerney}
\email{james\_mcnerney@hks.harvard.edu}
\affiliation{Center for International Development, Kennedy School of Government, Harvard University, Cambridge MA 02139, USA}
\affiliation{Complexity Science Hub Vienna, Vienna, Austria}

\author{Yang Li}
\affiliation{Center for International Development, Kennedy School of Government, Harvard University, Cambridge MA 02139, USA}

\author{Andres Gomez-Lievano}
\affiliation{Center for International Development, Kennedy School of Government, Harvard University, Cambridge MA 02139, USA}
\affiliation{Analysis Group, Inc., Boston MA 02199, USA}

\author{Frank Neffke}

\affiliation{Center for International Development, Kennedy School of Government, Harvard University, Cambridge MA 02139, USA}

\affiliation{Complexity Science Hub Vienna, Vienna, Austria}

\begin{abstract}
\noindent
\ed{Economic transformation -- change in what an economy produces -- is foundational to development and rising standards of living.  Our understanding of this process has been propelled recently by two branches of work in the field of economic complexity, one studying how economies diversify, the other how the complexity of an economy is expressed in the makeup of its output.  However, the connection between these branches is not well understood, nor how they relate to a classic understanding of structural transformation.}  Here, we present a simple \ed{dynamical modeling framework} that unifies these areas of work, based on the widespread observation that economies diversify preferentially into activities that are related to ones they do already. \ed{We show how stylized facts of long-run structural change, as well as complexity metrics, can both emerge naturally from this one observation.} However, \ed{complexity} metrics \ed{take on new meanings}, as descriptions of the long-term changes an economy experiences rather than measures of complexity per se.  \ed{This suggests relatedness and complexity metrics are connected, in a hitherto overlooked way: Both describe structural change, on different time scales.} \ed{W}hereas relatedness probes transformation on short time scales, complexity metrics capture long-term change.
\end{abstract}

\maketitle

\section{Introduction}
The prosperity of an economy is tied to the economic activities it can develop\cite{Hausmann2007}. Whereas places like Silicon Valley, the city of London, and the country of Japan pursue diverse and profitable activities, other places struggle to shift out of a narrow range of activities with low economic returns.  Working to understand why, the \ed{emerging} field of economic complexity has emphasized two branches of research.  In the first, researchers have asked how \ed{economies (e.g. countries, regions, cities) develop into new sectors of activity}.  One finding that \ed{repeatedly} emerges is that economic diversification typically entails shifts into activities that are related to ones that are already in place.  This tendency has been corroborated in research on the growth of industry clusters\cite{porter2003economic,delgado2014clusters}, and on related diversification by individuals \cite{gathmann2010general,guevara2016research}, firms\cite{farjoun1994beyond,lien2009using,neffke2013}, regions\cite{Neffke2011a,Boschma2012,boschma2013emergence,essletzbichler2015relatedness,zhu2017jump}, and countries \cite{hausmann2007structure,Hidalgo2007,Hausmann2021}. Recently these \ed{tendencies} have together been referred to as the \emph{Principle of Relatedness}\cite{Hidalgo2018}.

A second stream of research has \ed{investigated the concept of \emph{complexity}, a term that refers to the sophistication and diversity of the input needs of products, or of the productive endowments of places -- product complexity and place complexity, respectively.  More complex economies are expected to make more complex and more profitable products by providing a richer basis of productive capabilities.  A key objective of this stream of work has been to develop methods to infer the complexity of products and places} from data
\cite{Hidalgo2009,Tacchella2012,KempBenedict2014,Mariani2015,Morrison2017,Servedio2018,Teza2018,Schetter2019,Bustos2020,Sciarra2020,ivanova2020measuring,Teza2021,GomezLievano2021}.  This literature puts forward metrics to estimate complexity from the structure of \ed{a bipartite network that describes which locations produce which products in quantities suggesting a location has the needed capabilities to engage competitively in the product}.  Well-known complexity metrics include the Economic Complexity Metric (ECI) \cite{Hidalgo2009} and country Fitness \cite{Tacchella2012}, and others have been proposed in their wake \cite{Servedio2018,Teza2018,Bustos2020,Sciarra2020,Teza2021,GomezLievano2021}, \ed{such as GENEPY}.

While both branches of investigation have found success, the connection between them is not well understood\ed{, potentially limiting the development of this research agenda and its interpretability for scholarly and policy work.}  
Here, we present a simple \ed{modeling framework that links these branches}, and suggests they describe the same dynamical process of development, at different time scales and granularity. Our framework begins by making explicit the dynamics that are implicitly used in the first branch to describe economic diversification under the Principle of Relatedness (PoR).  We then follow a standard approach, analyzing these models with the workhorse method of eigenmode decomposition.  \ed{PoR} models describe economic structural change on short (e.g. year-to-year) time scales.  We work out the implications of these \ed{short-run descriptions} for economic evolution over the long run.  We show that the PoR implies the importance of at least two kinds of long-run changes in an economy's basket of activities.  One involves changes in the diversity of activities \ed{(such as the number of product categories in which an economy competitively exports). The} other involves changes in their relative mix or composition.  These changes are tracked by a pair of coordinates, one associated with diversity, the other with a particular pattern of shifts in an economy's activity basket.  

\ed{We then show (1) how these coordinates re-express a classic understanding of structural change, and (2) how they relate to complexity metrics.}  \ed{The two coordinates that emerge from the PoR} resonate surprisingly well with how economists have long described economic development in \ed{work} stretching back decades.  It is well known that countries diversify as they rise through lower and middle stages of income \cite{Imbs2003,Cadot2011}, and undergo compositional changes that include shifting out of labor-intensive forms of agriculture and into other sectors \cite{clark1940conditions,Kuznets1957}.  Motivated by this, we examine the coordinates that emerge from our framework in data on global production patterns, asking whether their movements capture these known stylized facts.  We use the coordinates generated by the \ed{framework} to describe 56 years of change in the export baskets of about 250 countries and regions.  What we find is that the dominant movements of these coordinates are intuitive and consistent with classic observations of economic development.  We \ed{see} simultaneous movements along the diversity and composition coordinates that correspond to a well-documented pattern of development: Countries diversify into making a greater number of products, while simultaneously shifting out of agricultural products toward manufactured goods (e.g. \cite{Imbs2003,Brummitt2020}).  This demonstrates the empirical relevance of our coordinates, and supports an interpretation of them as summary measures of the diversity and compositional changes associated with structural change.  This is also noteworthy because it shows that transient, year-to-year dynamics contain a great deal of information about the long-term, permanent development of economies.

\begin{figure*}[t]
\center
\includegraphics[width=0.63\textwidth]{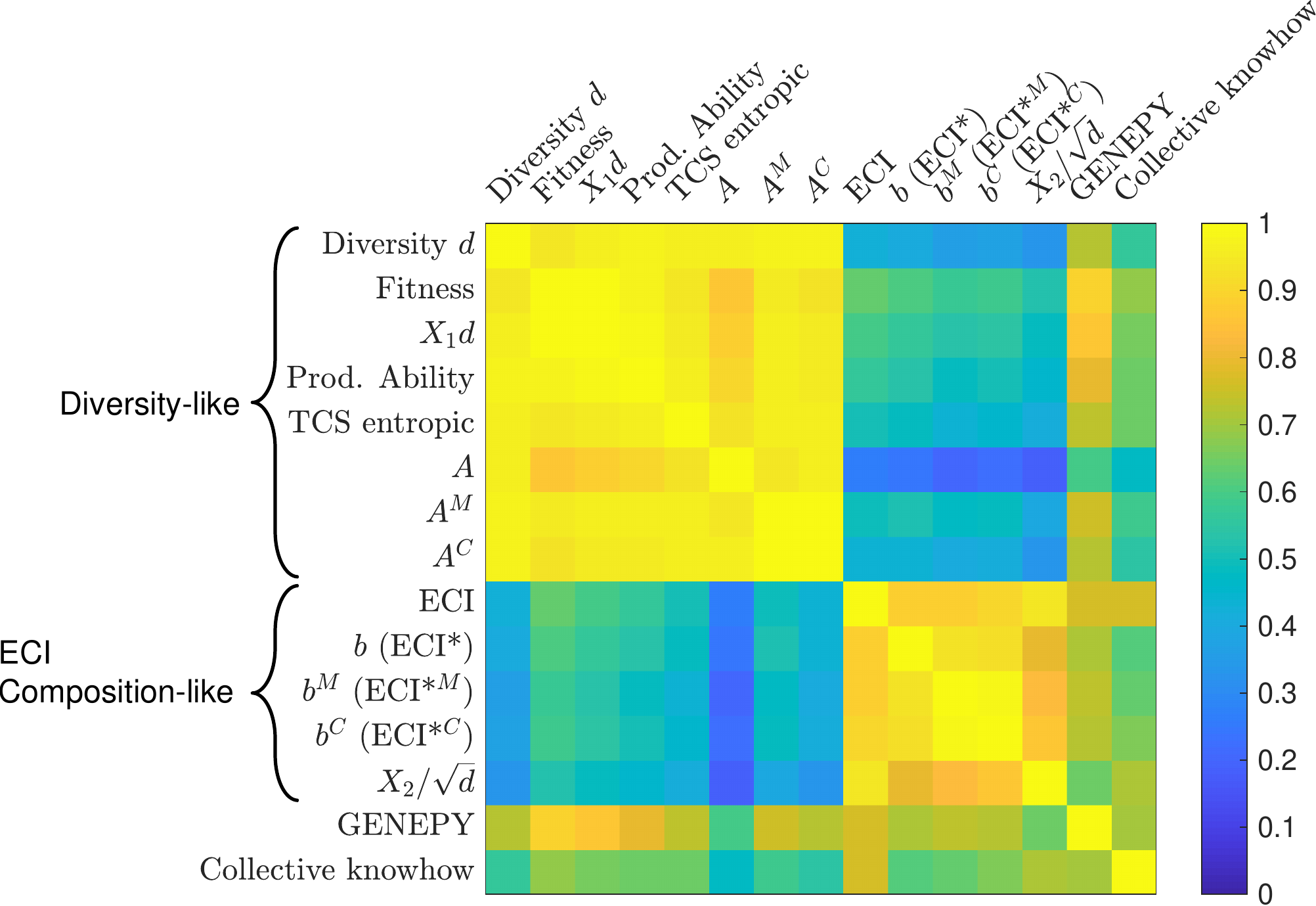}
\caption{\ed{Spearman rank correlations among complexity metrics and the coordinates that emerge from our dynamical modeling framework.  Diversity here is the count of activities in which a place has a revealed comparative advantage above 1, a common measure of the number of activities that a place performs competitively \cite{Hidalgo2009,Tacchella2012}, and often taken to have a close association with an economy's underlying complexity.  Proposed complexity metrics include the Economic Complexity Index (ECI)\cite{Hidalgo2009}, country Fitness\cite{Tacchella2012}, Production Ability\cite{Bustos2020}, GENEPY\cite{Sciarra2020}, Collective knowhow\cite{GomezLievano2021}, and the entropic measure of Teza, Caraglio, and Stella \cite{Teza2018} (here labelled TCS entropic).  $X_1$ and $X_2$ are the first and second components of the GENEPY metric \cite{Sciarra2020}.  $A$, $A^M$, and $A^C$ are different variations of the first coordinate that emerges from our dynamical framework, based on different literature approaches for estimating proximity between activities, and $b$, $b^M$, $b^C$ are the corresponding variations of the second coordinate that emerges from our framework. All complexity metrics and other quantities were computed for the year 2016 using UN Comtrade data \cite{Comtrade}.}}
\label{fig_complexityMetricCorrelations}
\end{figure*}

\ed{We then observe} a close correspondence between \ed{the} coordinates implied by the PoR and complexity metrics.  \ed{Nearly all complexity metrics fall in one of two groups, as shown in Fig. \ref{fig_complexityMetricCorrelations}}.  \ed{This is surprising given the variety of theoretical arguments that have been put forward to arrive at different metrics, and the emphasis on these arguments to justify one metric over another.} One group contains metrics that strongly correlate with an economy's diversity, and includes the country Fitness metric.  The other group captures compositional information about an economy, and includes the ECI.  These groups of complexity metrics correspond numerically and theoretically to the coordinates \ed{generated by our framework}.  That is, complexity metrics come in two \ed{main} types, and these types are ones that should be expected to describe long-run changes to an economy's basket of activities \ed{if short-run changes are accurately described by the PoR}.  \ed{Given this, we} propose a simple connection between the two main branches of work in economic complexity:  Complexity metrics describe the long-term changes to an economy's basket of activities that are implied by the Principle of Relatedness. In particular, the PoR delivers quantities that track an economy's \emph{diversity} and \emph{composition}.

\ed{In all, our modeling framework represents an approach to economic structural change that exploits dynamical systems methods, and in which the PoR and complexity metrics, as well as classic findings of structural change, are mutually reconciled.}  \ed{Besides the potential for using this framework as a starting point for further studies, many of our immediate findings relate to the understanding of complexity metrics that have become widely used in recent years \cite{Balland2022}.  We discuss these points in depth in section \ref{sec_complexity_metrics}.  Briefly, our framework grounds these metrics conceptually and mathematically on the PoR, a different basis from the heuristic arguments that have motivated these metrics, which ask how the complexity of a sector or economy can be inferred from data on what economies produce. Complexity metrics are typically seen as competing methodologies, each offering a potential solution to a challenging inference problem -- extracting country and product complexities from observable patterns of production.  But our results suggest the two main contending classes of metrics are not competing measures but give useful, complementary information about an economy's development.  Nevertheless, our results do not associate complexity metrics with complexity necessarily, and raise the question how well these metrics infer complexity per se versus summarizing long-run changes in an economy's basket of activities, becoming principled ways to restate and quantify classic statements of structural change.  Other more technical differences also arise from basing complexity metrics on the PoR, which we discuss in depth.}  Our paper also represents an effort to quantify structural change in economies, and thus it both joins recent literature, such as work focusing on regional and city development \cite{Neffke2011a,Boschma2012,boschma2013emergence,essletzbichler2015relatedness,zhu2017jump} or using new methods from machine learning \cite{Brummitt2020}, and provides continuity with \ed{classic literature in economics} \cite{clark1940conditions,Kuznets1957,Imbs2003}.

\begin{figure*}[t]
\center
\sidesubfloat[]{\includegraphics[width=0.6\textwidth]{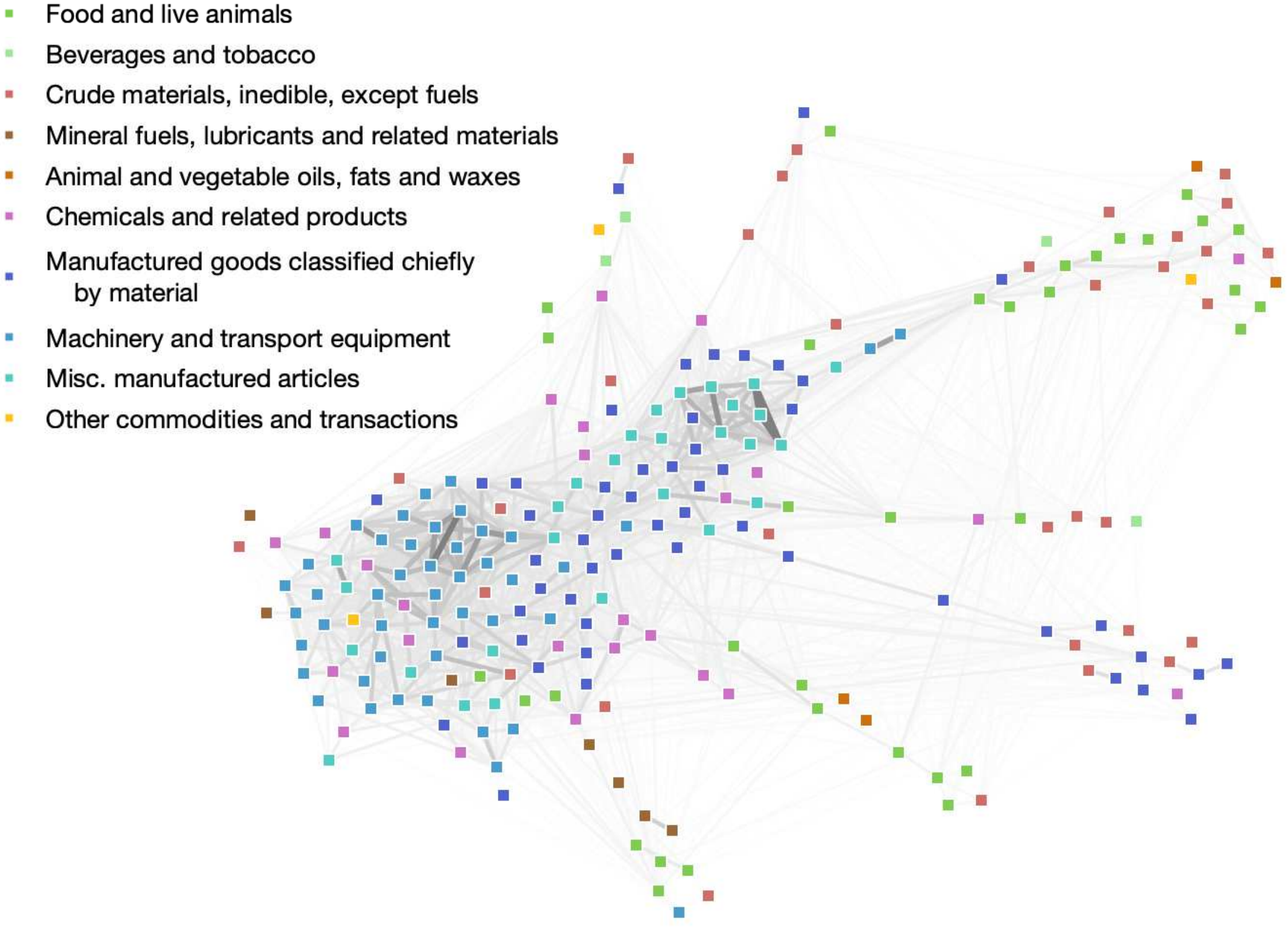}}\\
\sidesubfloat[]{\includegraphics[width=0.3\textwidth]{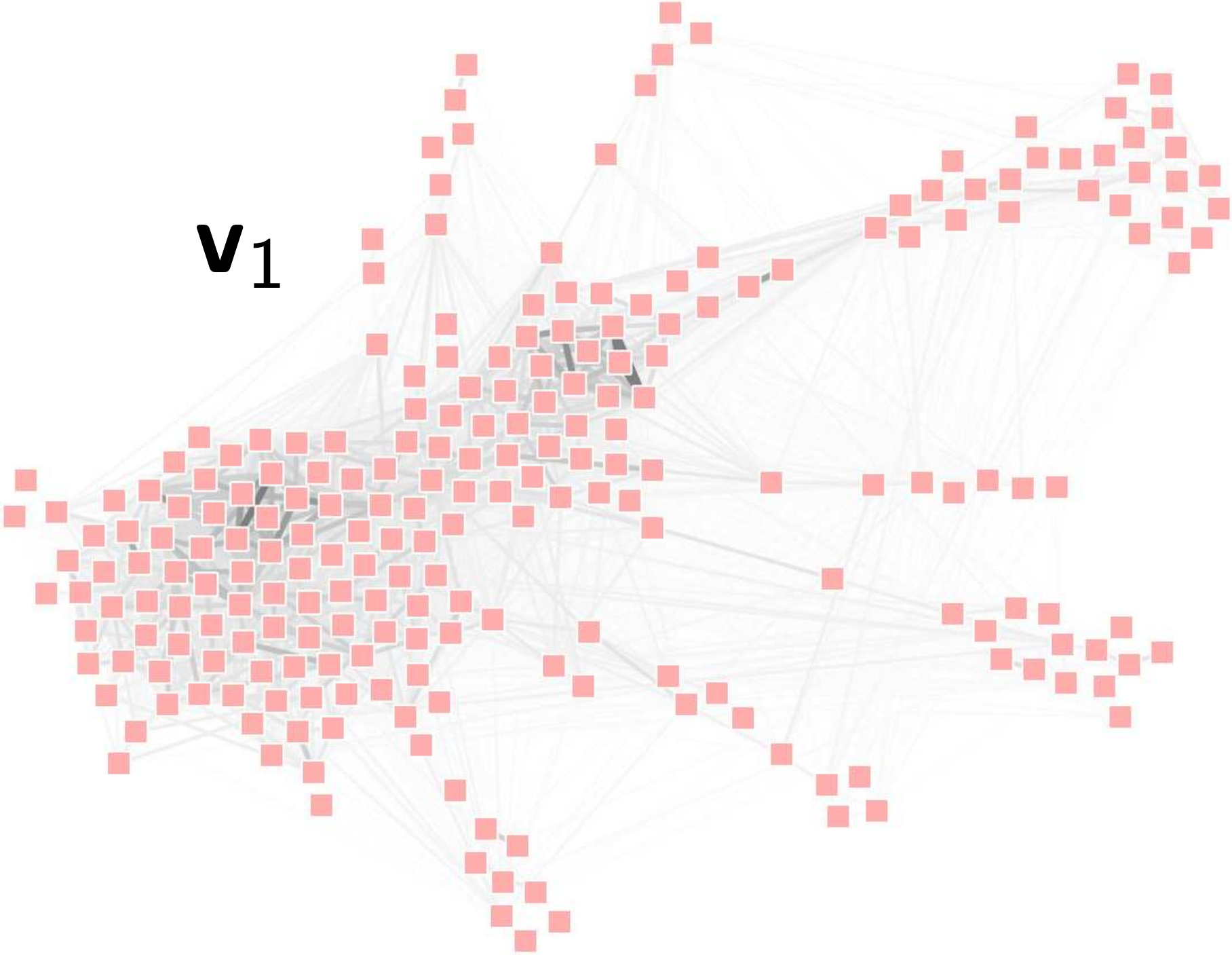}}
\sidesubfloat[]{\includegraphics[width=0.3\textwidth]{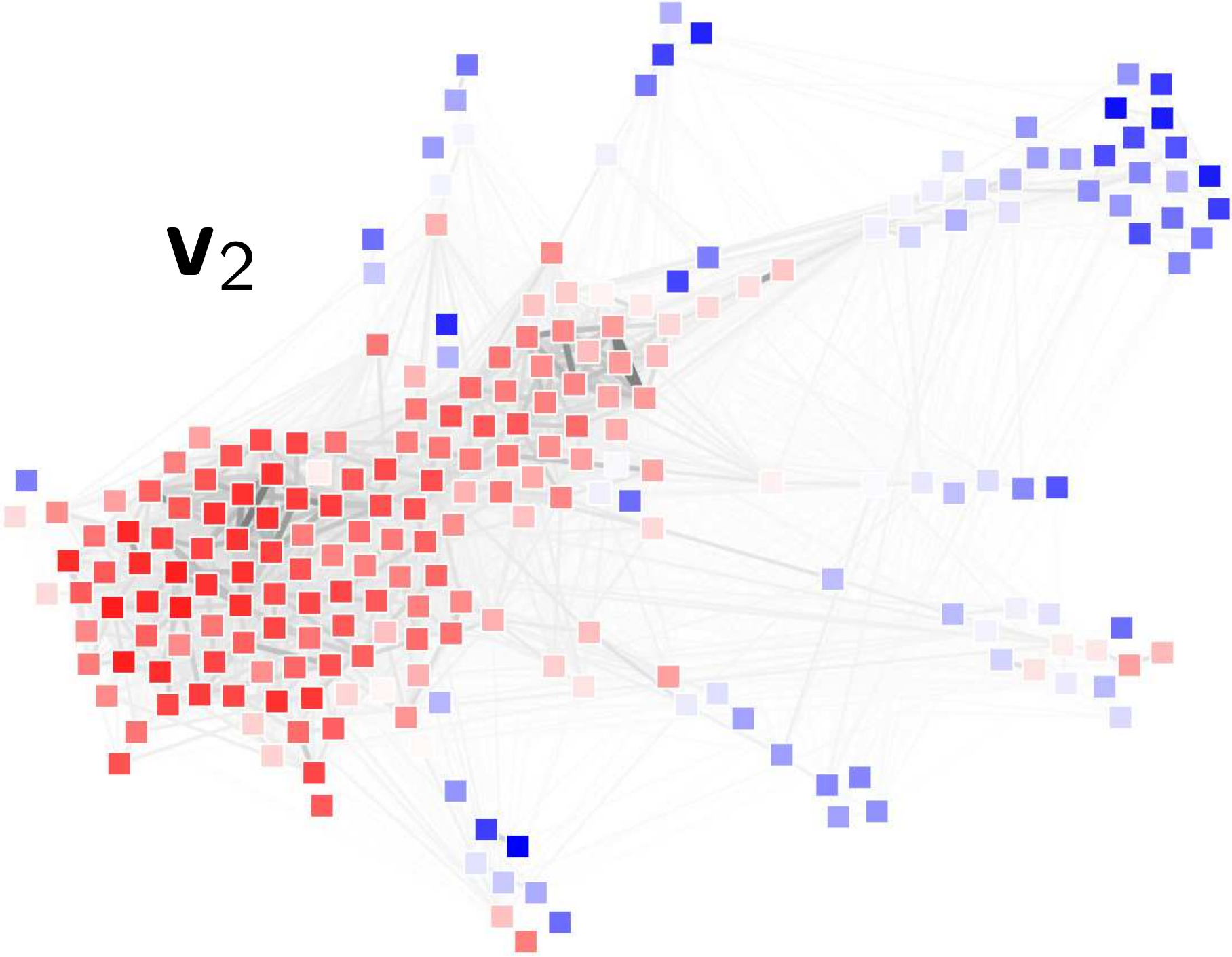}}
\caption{\ed{Network visualizations of relatedeness between tradable products. \textbf{a} To construct the network  above we use an approach similar to Hidalgo et al. (2007)\cite{Hidalgo2007}. Using UN Comtrade data \cite{Comtrade} we compute the revealed compared advantages (RCAs) of each country in each product $p$, and obtain the conditional probability across countries $Q_{pp'}$ that a country has an RCA greater than 1 in good $p$ given that it also meets this condition for $p'$.  We then compute relatedness between products $p$ and $p'$ as $\Phi_{pp'}^{M} \equiv \min( Q_{pp'}, Q_{p'p} )$.  For other approaches to constructing the network see e.g. Ref. \cite{Hausmann2021,vanDam2023}.}  \textbf{b-c} The first two right eigenvectors of \ed{the row-normalized proximity matrix} $\tilde{\Phi}_{pp'} = \Phi_{pp'} / \sum_{p''} \Phi_{p p''}$.  The first eigenvector is simply a uniform vector.  For the second eigenvector red nodes are positive entries while blue nodes are negative entries.}
\label{fig_product_space}
\end{figure*}

\section{Results}

\subsection{The PoR as a dynamical model}
A widespread finding about the geography of economic activity is that particular activities tend to coincide in the same places.  These patterns are often intuitive; a city, region, or country that makes cloth garments, for example, is also likely to make other textiles (e.g. knitted garments).  
These co-occurrences are assumed to indicate that two activities are related \ed{(i.e. they depend on similar underlying capabilities) and based on this various measures have been developed to infer the \emph{relatedness} of different activities.}
For concreteness, we focus on relatedness between exported products as inferred from \ed{their co-occurrences within} country export baskets.  \ed{Exports are often analyzed because harmonized data is available across countries of very different levels of development, and because exporting competitively represents an indicator of reaching an important level of production ability.}  Studies typically measure activity in an export using the Balassa index of revealed comparative advantage (RCA) \cite{Balassa1965}, $R_{cp} = \theta_{cp} / \theta_p$, where $\theta_{cp}$ is the share of country $c$'s exports devoted to product $p$, and $\theta_p$ is the share of $p$ in world exports.  RCAs are often treated as measures of inferred ability in exported products\cite{Hillman1980,Hinloopen2008}, and these quantities have also been used to characterize ability in activities besides exports (where the meanings of $\theta_{cp}$ and $\theta_p$ are adjusted appropriately).  Here we follow suit, though it is not an intrinsic requirement of our model to measure ability this way.  

The relatedness, also called proximity, of products $p$ and $p'$ is then computed by a measure of similarity between the $p$ and $p'$ columns of the matrix $R = [R_{cp}]$.  This means that $p$ and $p'$ are taken to be closely related if high ability in $p$ co-occurs with high ability in $p'$ across locations.  As one example, Hidalgo et al.  (2007)\cite{Hidalgo2007} say a country has significant ability in a product if its RCA in the product exceeds 1, and compute the conditional probability $Q_{pp'}$ that country $c$ has an RCA greater than 1 in good $p$ given that it also satisfies this condition for $p'$.  The proximity $\Phi_{pp'}$ between products $p$ and $p'$ is taken to be the lesser of $Q_{pp'}$ and $Q_{p'p}$: $\Phi_{pp'}^{M} \equiv \min( Q_{pp'}, Q_{p'p} )$.  The structure of proximities between products is often \ed{visualized} using network representations (Fig. \ref{fig_product_space}\ed{a}).

Crucially, changes in an economy's activities are predicted by the structure of proximities. Empirical studies show that the development of high ability in an exported product tends to be preceded by high ability in nearby products \cite{Hidalgo2007}.  A typical \ed{regression} modeling setup \cite{boschma2013emergence,boschma2015relatedness,zhu2017jump,petralia2017climbing} to explore this effect takes future ability in a product $p$ to be a function of the \emph{density} around it, defined as the average ability that an economy has in other products, weighted by proximity to $p$:
\begin{align}
\ed{R_p(t+1) = b_1 R_{p}(t) + b_2 \sum_{p'} \frac{\Phi_{pp'}}{\sum_{p''} \Phi_{pp''}} R_{p'}(t) + \varepsilon_{p}(t)}.
\label{eq_densityRegression}
\end{align}
An alternate setup \cite{Hausmann2021} contains two steps, with ability in product $p$ first regressed on density in the same time period, establishing the existence of systematic correlations in which economic activities co-occur.  Residuals from this regression are then used to forecast appearances of comparative advantage in future periods.  Combining two such regressions analytically will also lead to an expression of the form of Eq. \eqref{eq_densityRegression}.  These types of analyses overwhelmingly find a robust, positive statistical association between density and future growth and diversification, bolstering the idea that relatedness measures capture underlying similarities in activities that make some transitions easier to achieve than others, influencing the direction in which an economy develops.

\ed{In general, t}he PoR treats diversification as a process of spreading on a network \ed{of economic activities}. \ed{One such example is the network of products in Fig. \ref{fig_product_space}.}  \ed{The relatedness network is typically taken to be fixed for the purpose of predicting this process.}  Using Eq. \eqref{eq_densityRegression} we can make make more explicit the dynamical model the PoR implies.  \ed{The results that follow use network concepts (see e.g. \cite{Newman2010}) and the dynamical systems method of eigenmode decomposition (see e.g. \cite{Schaub2019}).}  A given economy has various levels of ability in different activities, which are given by elements of a vector $\vec{R}(t)$.  We call this the economy's activity basket.  These abilities evolve as the economy improves or shifts into new parts of the network.  Using Eq. \eqref{eq_densityRegression} as a guide, a simple model of the diversification process is
\begin{align}
\dot{\vec{R}}(t) = 
\gamma(t) \vec{R}(t) - \frac{1}{\tau} L_\Phi \vec{R}(t),
\label{eq_diffEq}
\end{align}
where $\gamma(t)$ is an arbitrary growth rate, $L_\Phi$ is a graph Laplacian, and $\tau$ is a time scale.  The two terms on the right capture two kinds of changes.  An economy's abilities can rise or fall as a whole, and they can shift according to the Laplacian term to weight activities differently, changing which ones receive the most emphasis.
 
The Laplacian term \ed{reflects} an implicit choice of empirical specifications of the PoR as instances of consensus dynamics (e.g. \cite{Schaub2019}).  The Laplacian $L_\Phi$ captures the structure of the network and governs shifts between activities, taking the form $L_\Phi = I - \tilde{\Phi}$ where $\tilde{\Phi}$ is a proximity matrix with rows normalized to sum to 1, $\tilde{\Phi}_{pp'} = \Phi_{pp'} / \sum_{p''} \Phi_{p p''}$.  With this assumption, a discrete-time approximation of Eq. \eqref{eq_diffEq} in index form reads
\begin{align}
&\frac{R_p(t+\Delta t) - R_p(t)}{\Delta t} \approx \nonumber\\
&\quad \gamma(t) R_p(t) + \frac{1}{\tau} \left( \sum_{p'}\tilde{\Phi}_{pp'} R_{p'}(t) - R_p(t) \right).
\label{eq_discreteApproximation}
\end{align}
In this form the dynamical model is easily compared with regression setups that test for the PoR in the literature.  Ability in activity $p$ changes between periods in a way that depends on ability in nearby products.  When the average ability in activities around $p$ ($\sum_{p'}\tilde{\Phi}_{pp'} R_{p'}(t)$) exceeds that in $p$ itself ($R_p(t)$), ability in $p$ rises.  \ed{The last term in Eq. \eqref{eq_discreteApproximation} can also be negative, corresponding to a decline in ability.}  In addition, abilities have the freedom to rise or fall as a whole because of the term $\gamma(t) R_p(t)$.

Integrating Eq. \eqref{eq_diffEq} over time leads to
\begin{align}
\vec{R}(t) = A(t) P(t) \vec{R}(0),
\label{eq_solution}
\end{align}
where the dynamics are now captured by the scalar $A(t)$ and the matrix $P(t)$.  The prefactor $A(t) \equiv e^{\int_0^t \gamma(s) \;ds}$ is a shift factor that captures accumulated growth in overall ability up to time $t$, scaling the $\vec{R}$ vector up or down as a whole.  The matrix $P(t) = e^{-L_\Phi (t/\tau)}$ is a stochastic matrix that transforms $\vec{R}$, changing the relative emphasis on different activities over time.

Implementing this model involves some practical considerations. First, RCAs are commonly transformed non-linearly to weaken the influence of extreme values \cite{hoen2006measurement,elekes2019foreign,Brummitt2020,Hausmann2021}. When we use RCAs as a measure of ability, we will do so as well, but note that this does not change the model in any fundamental way. Second, we need to choose how to operationalize the network of relatedness $\Phi_{pp'}$ between activities.  Studies have used a variety of measures of proximity.  We consider several options, finding similar outcomes, as we discuss later.

\subsection{A PoR-derived coordinate system to describe an economy's activity basket}
Dynamical models are frequently analyzed \ed{in terms of} their dynamical modes, called eigenmodes in linear models. Applying eigenmode decomposition to the model above leads to a coordinate system that can be used to describe the evolution of an economy's activity basket.  Let $\vec{v}_\mu$ be the $\mu$th right eigenvector of $L_\Phi$ and let $\kappa_\mu$ be its eigenvalue.  Ordering eigenvalues from least to greatest, the first right eigenvector has $\kappa_1 = 0$ and is a uniform vector of real positive numbers.  It can be taken to be a vector of 1s, $\vec{v}_1 = \vec{1}$.  Higher-order eigenvectors ($\vec{v}_2$ and up) have $\kappa_\mu > 0$, and contain a mix of elements with positive and negative real parts.%
\footnote{Depending on the proximity matrix used, some eigenvectors could contain imaginary parts, but this would pose no special difficulties in interpreting modes; see \ed{Supplementary Note S1}.} %
The right eigenvectors of $L_\Phi$ form a basis for the vector space of activity baskets $\vec{R}(t)$, and as a result, one can write any such vector as a linear combination $\vec{R}(t) = \sum_{\mu} c_{\mu}(t) \vec{v}_\mu$.  Letting $\vec{w}_\mu$ be the $\mu$th \emph{left} eigenvector of $L_\Phi$, the coefficients $c_{\mu}(t)$ may be computed by exploiting the biorthogonality of left and right eigenvectors, $\vec{w}_\mu^\dag \vec{v}_\nu = \delta_{\mu\nu}$\ed{, giving} $c_{\mu}(t) = \vec{w}_\mu^\dag \vec{R}(t)$.

Following the usual steps of eigenmode decomposition, one can decompose $\vec{R}(t)$ to separate different modes of change according to their time scales.  \ed{Plugging $\vec{R}(0) = \sum_{\mu} c_{\mu}(0) \vec{v}_\mu$ into Eq. \eqref{eq_solution} gives $\vec{R}(t) = A(t) \sum_\mu c_{\mu}(0) e^{-\kappa_\mu t /\tau} \vec{v}_\mu$.  The first term contains the first eigenvector $\vec{v}_1 = \vec{1}$ and corresponds to the fixed point of the simpler model $\vec{R}(t) = P(t) \vec{R}(0)$ (i.e. the model with the factor $A(t)$ fixed at 1).  Separating the first term from other terms of the sum we have}
\begin{align}
\vec{R}(t) = A(t) \left( c_1(0) \vec{1} + \sum_{\mu\geq 2} c_{\mu}(0) e^{-t/\tau_\mu} \vec{v}_\mu \right),
\label{eq_eigenmodeDecomposition}
\end{align}
where we defined $\tau_\mu \equiv \tau / \kappa_\mu$.

Eq. \eqref{eq_eigenmodeDecomposition} describes the following behavior.  First, neglecting the effect of the shift factor $A(t)$, the activity basket of a region converges over time to a uniform vector $c_1(0) \vec{1}$.  In this state, the region has equal ability in all products.  On its way to this state, the basket shows higher ability in some products and lower ability in others, and each eigenvector $\vec{v}_\mu$ describes a different pattern of deviations from the long-term steady state in which all activities are equally important.  Because each eigenvector $\vec{v}_\mu$ (for $\mu \geq 2$) satisfies $\vec{w}_1^\dag \vec{v}_\mu = 0$, where $\vec{w}_1$ has only positive elements, each such eigenvector has some elements that are positive and others that are negative.  Each of these patterns of deviation decays with time at the rate set by $\tau_\mu$, the characteristic time scale of the $\mu$th eigenmode.  If the pre-factor $A(t)$ were fixed and equal to 1, then a region would be destined to have ability $c_1(0)$ in every product.  Letting $A(t)$ change over time, the final level of ability in products can be arbitrarily high or low.


To better understand the implications of the PoR for economic change over the long term, we focus on the two modes of change associated with dynamics on the longest time scales.  Absorbing the coefficient $c_1(0)$ into $A(t)$, and defining the coefficient $b(t) \equiv c_2(0) e^{-t/\tau_2}$, the longest-lived dynamics of $\vec{R}(t)$ are described by the two leading terms of Eq. \eqref{eq_eigenmodeDecomposition}:
\begin{align}
\vec{R}(t) 
&\sim A(t) \vec{1} + A(t) b(t)\vec{v}_2.
\label{eq_bigR_eigenmode decomposition}
\end{align}
The coefficients $A(t)$ and $b(t)$ have simple interpretations.  A shift in $A(t)$ corresponds to a region realizing a uniform change in abilities across activities.  A shift in $b(t)$ corresponds to a compositional shift.  Some activities rise in ability and others fall, as determined by the signs and magnitudes of entries in $\vec{v}_2$.  Together, these coordinates situate the activity basket $\vec{R}(t)$ in a 2D space.  

To further interpret the $A(t)$ coordinate, let $\vec{\pi} = \vec{w}_1$ denote the first left eigenvector of $L_\Phi$, normalized so that its elements sum to 1, and note that $\bar{R}(t) = \vec{\pi}^T \vec{R}(t)$ defines a weighted average of the abilities of a region.  By again exploiting the biorthogonality of left and right eigenvectors, it can be shown (\ed{see Methods, Eq. \eqref{eq_Acoordinate_as_average}}) that $A(t)$ equals this average:
\begin{align}
A(t) = \bar{R}(t) \equiv \vec{\pi}^T \vec{R}(t).
\end{align}
For this reason, we refer to $A(t)$ as the average ability coordinate.

To further interpret the $b(t)$ coordinate, note that the activity basket $\vec{R}(t)$ conveys two types of information: the relative mix or composition of activities, as well as ability levels.  
Factoring out an economy's average ability $A(t)$ from $\vec{R}(t)$ gives a normalized vector, $\vec{r}(t) \equiv \vec{R}(t) / A(t)$, that characterizes only the composition of activities.  In particular, dividing Eq. \eqref{eq_bigR_eigenmode decomposition} by $A(t)$ we see that $\vec{r}(t) \sim \vec{1} + b(t) \vec{v}_2$, showing that $b(t)$ characterizes the deviation of the compositional vector $\vec{r}(t)$ from uniformity.

\begin{figure*}[t!]
\center
\sidesubfloat[]{\includegraphics[height=0.25\textwidth]{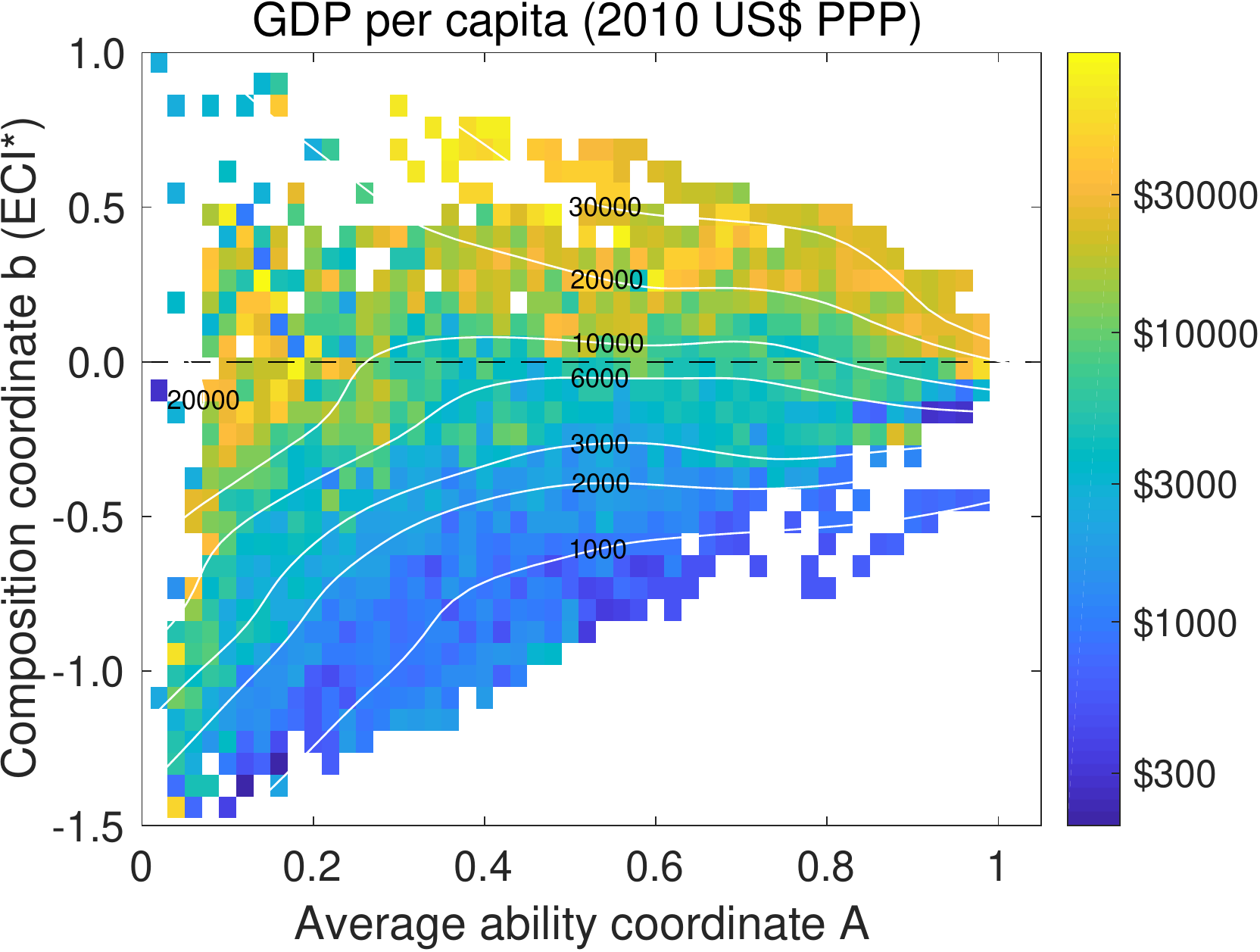}}
\hfill
\sidesubfloat[]{\includegraphics[height=0.25\textwidth]{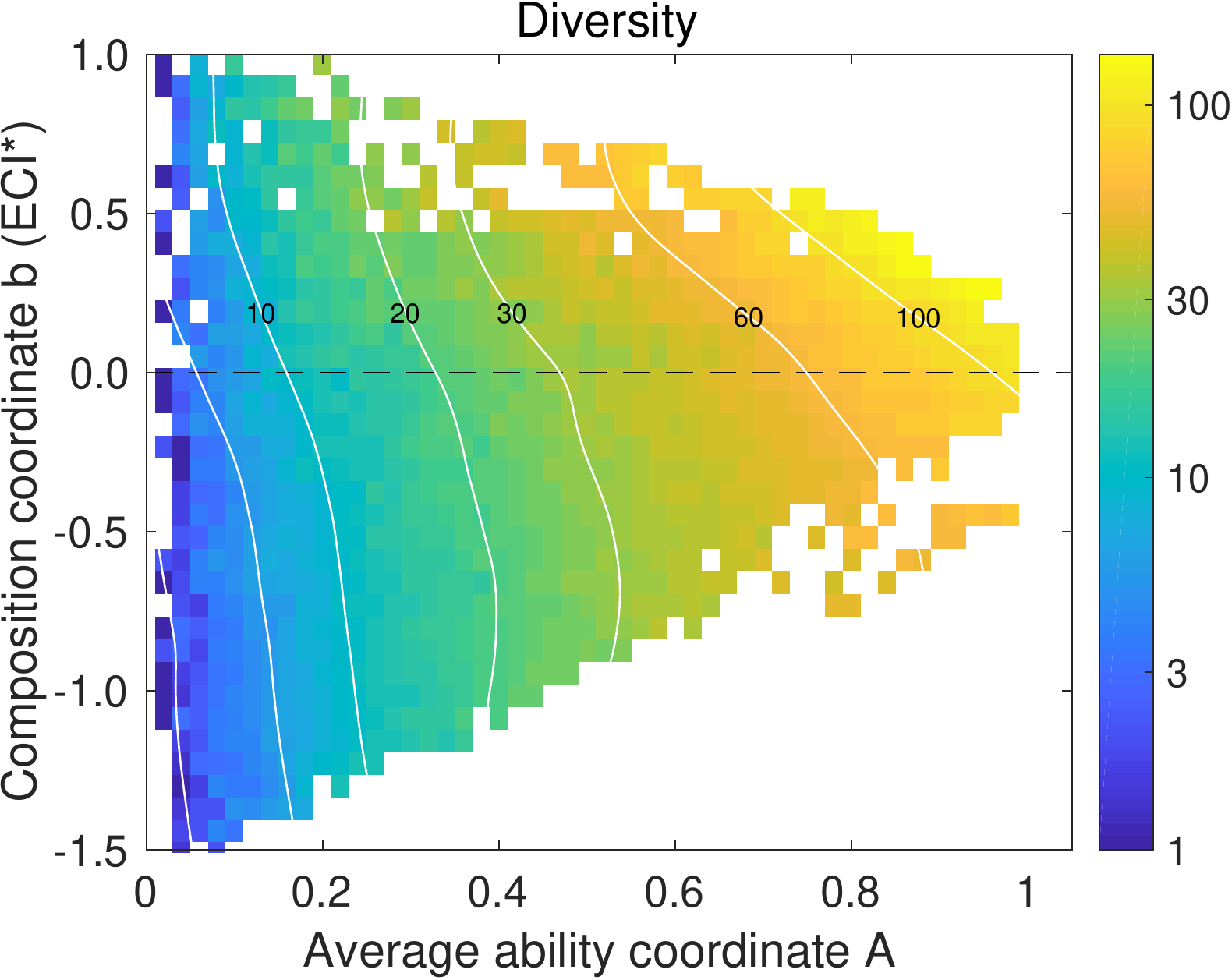}}
\hfill
\sidesubfloat[]{\includegraphics[height=0.24\textwidth]{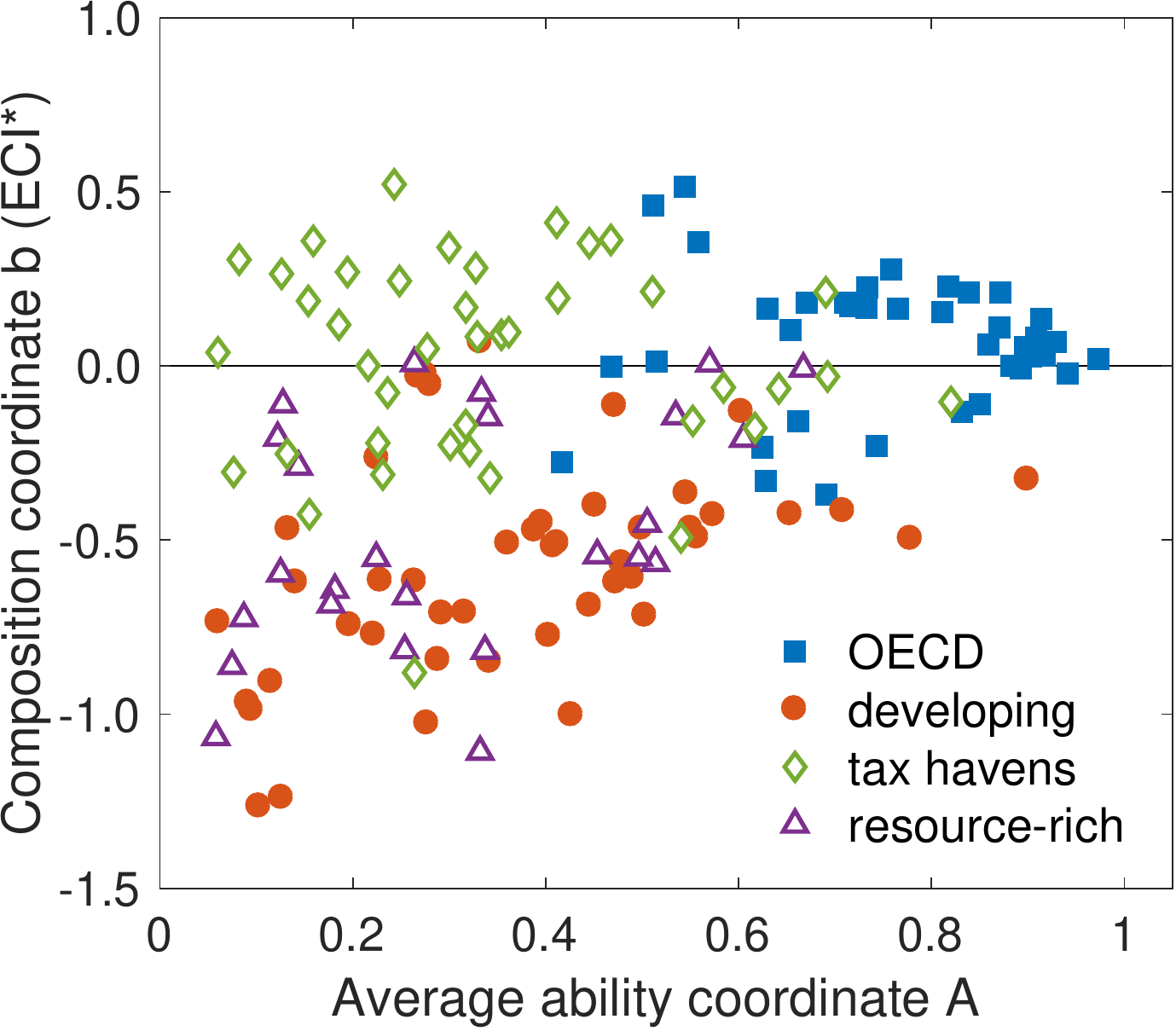}}
\vspace{10pt}
\sidesubfloat[]{\includegraphics[height=0.25\textwidth]{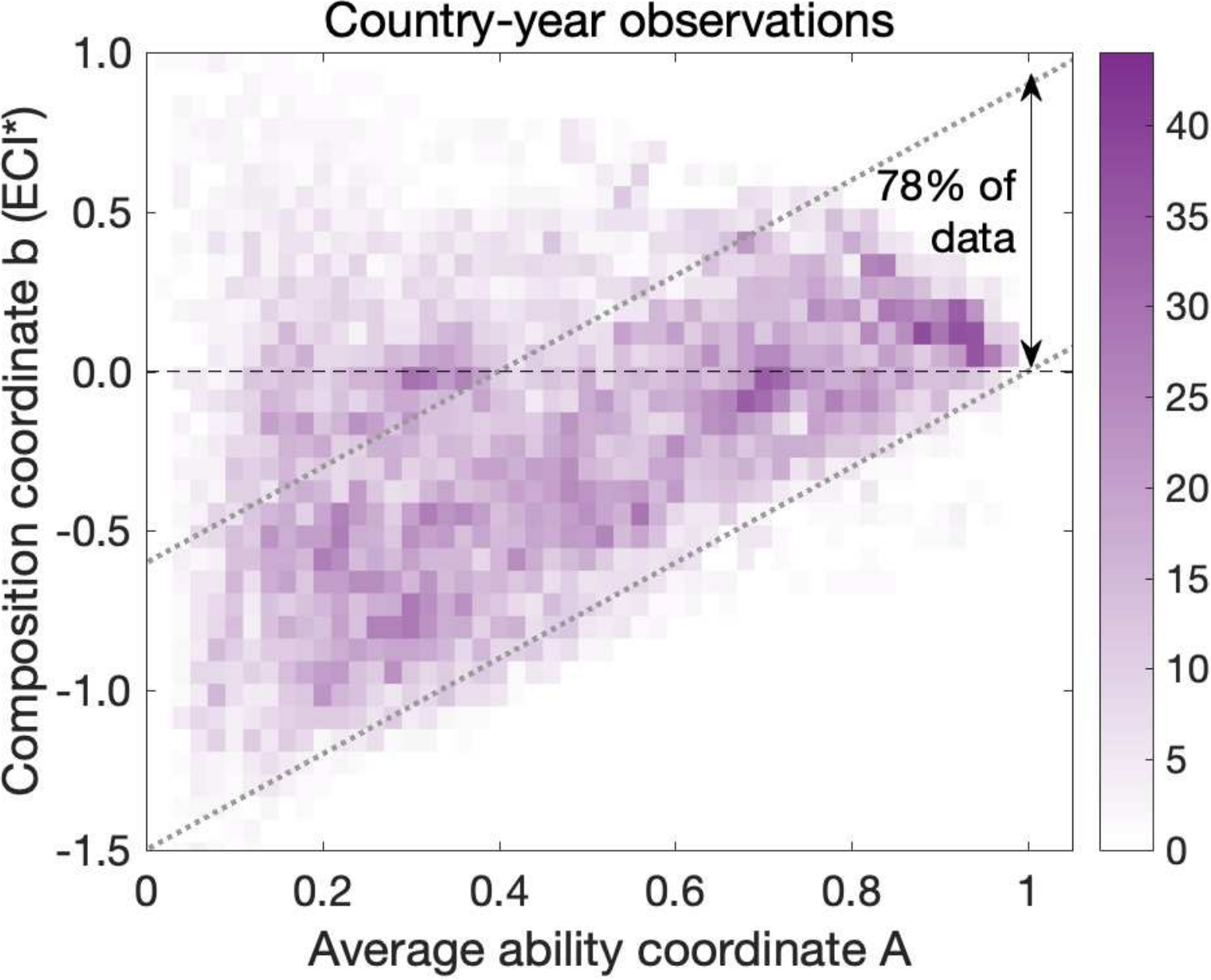}}
\hfil
\sidesubfloat[]{\includegraphics[height=0.24\textwidth]{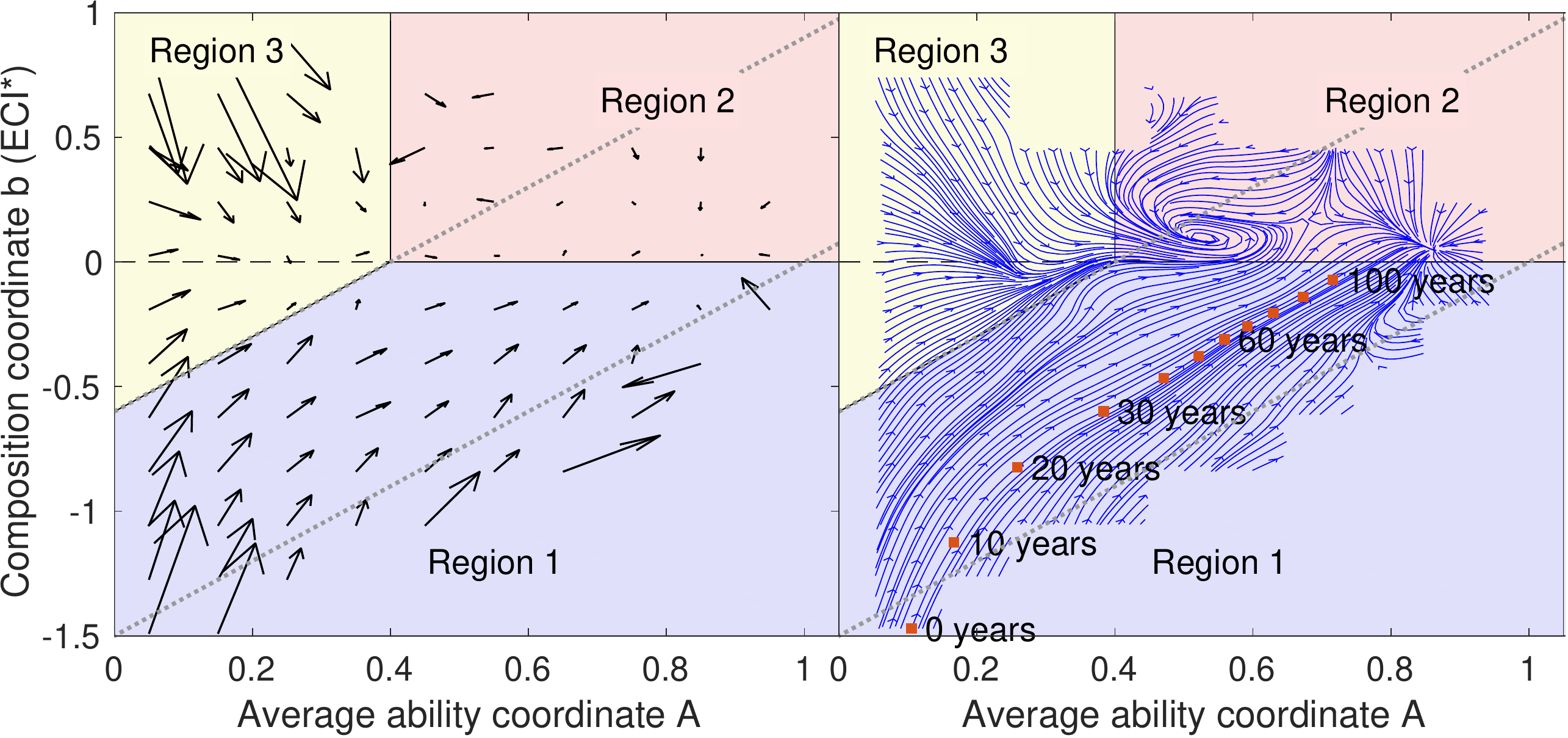}}
\caption{Country export baskets plotted according to average ability ($A$ coordinate) and composition ($b$ coordinate), based on Comtrade data \cite{Comtrade} for 249 countries and regions over the period 1962 - 2018 (11,544 observations).  \ed{We analyze these data at the 3-digit SITC product level (235 product categories).} \textbf{a}-\textbf{b} We form bins in the 2D plane to collect country-year observations, and compute average GDP per capita and average diversity $d$ within each bin. White lines are contours of a LOWESS-smoothed surface fit to bin averages.  \textbf{c} Positions of export baskets in 2018 of four categories of economies.  `Developing' refers to economies on the UN's least developed economies list \cite{UN_LDC}, `tax havens' to non-OECD and non-developing countries appearing in Hines (2010) \cite{hines2010treasure}, and `resource-rich' to economies not elsewhere categorized whose natural resource rent\cite{WorldBank} exceeded 5\% of GDP in 2018. \textbf{d} Number of country-year observations in each bin. \textbf{e}  Quiver and streamline plots for the average directional change over 20 years for countries beginning at different starting points in the plane.  Red dots in the right panel follow one streamline and roughly convey the average pace of movement, though no streamline should be taken as a typical country trajectory; individual countries display wide variation in both direction and speed of change. \ed{To construct the quiver plot, we divided the plane into 10 equal-size bins horizontally and 12 equal-size bins vertically (120 bins total).  For each bin we obtain the horizontal and vertical components of movement by computing the average horizontal change $\Delta A$ and average vertical change $\Delta b$ over the next 20 years using all observations starting in the bin.  We used the MATLAB quiver function to render a vector field and the streamslice function to render a streamline plot.}}
\label{fig_phasePlane}
\end{figure*}

\subsection{Diversity and composition of exports baskets over time}
The PoR thus delivers coordinates that theoretically could be used to characterize the activity basket of an economy.  We now examine these coordinates empirically, asking whether their movements corroborate known facts of economic development.  We focus on international trade data from UN Comtrade \cite{Comtrade}, which are frequently used in research on economic complexity, because they offer a detailed description of what economies export over long periods of time. This allows us to construct our coordinates to describe the evolution of countries' export baskets from 249 countries over the 57-year period 1962 - 2018.  This exercise summarizes country development by two simple statistics.

RCAs are commonly transformed to mitigate the influence of extreme values, and here we transform the heavy-tailed RCAs with the function $g(R) \sim \log(1 + R/R_0)$.  We tune the parameter $R_0$ such that it maximizes the variance that the $b$ coordinate can explain across time and countries \ed{(Methods section, ``Transformation of RCAs'').} We then calculate for each country $c$ and time $t$ the RCA vector $\vec{R}_c(t)$ on this transformed scale and use these vectors to compute the $A$ and $b$ coordinates.

Figs. \ref{fig_phasePlane}a-b depict country income and export diversity as functions of $A$ and $b$. For reasons that will become clear in the next section, we also refer to $b$ as ECI* in \ed{these graphs}.  Plotting income and diversity this way allows us to ask how each of these variables vary as a function of one coordinate while holding the other fixed.

\paragraph{Income increases with $b$, diversity with $A$.}  
A higher value of the compositional coordinate $b$ for a country's export basket is associated with significantly higher GDP per capita (Fig. \ref{fig_phasePlane}a).  For example, holding $A$ fixed at 0.5, an increase in $b$ from -0.5 to +0.5 is associated with an increase from about \$1500 to \$30,000 (2010 US dollars PPP).  In contrast, the association between a country's average ability $A$ and its income when holding the value of $b$ fixed is weak at best.

The situation is roughly opposite when we examine countries' economic diversity in these coordinates (Fig. \ref{fig_phasePlane}b).  The diversity of a country's economic activity has been quantified with a number of measures, such as the Gini coefficient and Herfindahl-Hirschmann index (e.g. \cite{Imbs2003}), or the count of products $d_c$ in which a region $c$ has an RCA greater than 1.  Higher values of the average ability coordinate $A$ are closely associated with greater diversity in a country's export basket.  Holding $b$ fixed at 0, an increase in $A$ from 0.1 to 0.9 is associated with an over 10-fold rise in \ed{the} number of products with RCA above 1.  In contrast, when holding $A$ fixed, a compositional shift towards higher levels of $b$ is only weakly associated with higher export diversity.

Different types of countries inhabit different regions of this coordinate system (Fig. \ref{fig_phasePlane}c). The industrialized countries of the OECD are predominantly located in the right of the plot with relatively high values of both $A$ and $b$, corresponding to diverse, high-income, developed economies.  In the lower left we find a set of undiversified developing economies as defined by the UN's least developed economies classification.  In contrast, in the upper left, we find countries that are also undiversified, yet \ed{frequently} have high-income.  Many of these are resource-rich economies or are often considered tax havens \cite{hines2010treasure}.

Note that, although countries often diversify as they rise in income\cite{Imbs2003}, the observations above do not directly associate higher income with higher diversity of exported products.  Rather, they suggest that what matters more for a country's GDP per capita  is the composition of exports\cite{Hausmann2007}.  We next explore the dynamics of countries in this space and probe this distinction further.

\paragraph{Country development.}
Export baskets occupy a triangular region of the $A$-$b$ phase plane.  A particularly densely occupied portion of this plane is a diagonal band that stretches from low $A$ and low $b$, to high $A$ and above-average $b$ (Fig. \ref{fig_phasePlane}d).  We start by focusing on the movement of countries whose export baskets lie in this band.  Countries show diverse trajectories (\ed{see Supplementary Note S2}).  To nevertheless characterize broad tendencies over long periods, we group country-year observations into bins in the $A$-$b$ plane.  We examine the average direction and speed of movement over the next 20 years for observations that \ed{start} within a given bin.  For expositional convenience, we summarize our results by referring to three regions of the plane, labelled Regions 1, 2, and 3 (Fig. \ref{fig_phasePlane}e).

Countries beginning in the lower part of the main diagonal band -- i.e. in Region 1 -- trend over decadal time scales in a lower-left-to-upper-right direction.  Given how the coordinates $A$ and $b$ are defined, a simultaneous increase in both directions corresponds to an export basket that simultaneously realizes two kinds of changes:  (1) a general improvement in ability across products; (2) a shift in composition towards products with positive values in the vector $\vec{v}_2$.  These two components of movement are associated with different effects.  The shift in the composition of exports captured by increasing $b$ is (empirically) most directly associated with increased GDP per capita (Fig. \ref{fig_phasePlane}a).  In contrast, the increase in $A$ is more associated with a rise in export diversity  (Fig. \ref{fig_phasePlane}b).  These patterns are consistent with the idea that higher \ed{income} is associated with exporting particular products, rather than diversification per se.  Nevertheless, in practice, countries in the lower part of the main diagonal band that succeed in reaching such products tend do so while simultaneously diversifying into a broad range of goods, including ones not associated with higher \ed{income}.

What products are involved in the vertical shifts that increase $b$? Manufacturing products very often have positive elements in $\vec{v}_2$, and agricultural products very often have negative elements.  As a result, countries in Region 1 that traverse the length of the diagonal band see a broad, long-lived shift in their export baskets away from agricultural products and towards manufactured products. This shift is consistent with the long-observed tendency for economies to move from agriculture to manufacturing (and then on to services, a move that largely eludes trade statistics) as they develop\cite{clark1940conditions,Kuznets1957}.

If a country traverses the length of the diagonal band in Region 1 it will arrive in Region 2, where countries have high-income, diverse, developed economies.  Within this region, the average speed of movement is much lower than in Region 1. This in part owes to the fact that export baskets here evolve in a greater variety of directions, with an average directional change near zero.  Broadly though, countries in Region 2 tend to sustain a high $b$, somewhat above zero, and move within a range of relatively high $A$ values.  On reaching Region 2, a number of countries move toward lower values of $A$.  This transformation path is consistent with a phenomenon in which countries see a fall in diversity in late stages of development\cite{Imbs2003,Cadot2011}. 

Finally, countries in Region 3 tend to have high income and low diversity.  Many of these countries are abundant in natural resources (particularly oil) or function as tax havens.  In this region, countries tend to move quickly toward lower levels of $b$, converging near zero.  We are not aware of any prior observations that this movement corresponds to.  Evidently it is difficult for these countries to sustain a high degree of specialization in products that load positively on $\vec{v}_2$ for long periods of time.

Together, these observations show an end-to-end agreement of our framework between short-term and long-term changes in economic development.  The Principle of Relatedness describes structural change on short time-scales, and fine-grained levels of sectoral resolution.  When we analyze network models that operationalize this principle, using the standard technique of eigenmode decomposition, we arrive at coordinates that should capture activity changes over long time-scales, and higher levels of aggregation.  As one would hope, observed movements of export baskets in these coordinates yield long-term and coarse-grained descriptions of structural change that are consistent with well-documented stylized facts.

\ed{Before moving on, we note that higher-order eigenvectors beyond the second are of interest because in principle they could describe other important modes of transformation.  However, the proximity matrices in the literature only show strong agreement in the structure of the first two eigenvectors, and do not straightforwardly resolve how many eigenvectors matter (see Supplementary Note S3).}%

\subsection{Comparing our structural change coordinates with complexity metrics}\label{sec_complexity_metrics}
We see that the PoR leads naturally to coordinates that track the process of structural change.  We now show that the resulting coordinates closely resemble complexity metrics that have been proposed in recent years, even though the latter have been motivated along very different lines than the coordinates we derive here.  Complexity metrics have been put forward as practical tools to draw inferences about the number of distinct production capabilities that different economies possess, based on observations about which activities are performed in which places.  Among the metrics that have been proposed are the Economic Complexity Index (ECI) \cite{Hidalgo2009}, country Fitness \cite{Tacchella2012, Servedio2018}, the entropic measure of Teza, Caraglio, and Stella \cite{Teza2018,Teza2021}, Production Ability \cite{Bustos2020}, GENEPY \cite{Sciarra2020}, and collective knowhow \cite{GomezLievano2021}.  We first describe the relationships between these metrics and our coordinates and then comment on their significance afterwards.

Different complexity metrics are set apart by many differences in motivating narrative and implementation.  Despite this, empirically, complexity metrics fall into two main groups that emphasize different kinds of information (Fig. \ref{fig_complexityMetricCorrelations}).  These groups have a straightforward correspondence with the coordinates generated from the PoR.  The first group contains diversity-like quantities.  This includes diversity $d_c$ itself, the $A$ coordinate of our \ed{framework}, country Fitness, Production Ability, and the entropic measure of Teza, Caraglio, and Stella.  The other group contains quantities that capture a particular type of compositional information about an economy; roughly, the agriculture-to-manufacturing axis noted earlier.  This group includes the ECI, and the $b$ coordinate of our \ed{framework} (ECI*).  The GENEPY metric by design combines information associated with both groups, and itself does not fall clearly in either one, but its first component $X_1$ by construction is related to Fitness and is associated with the first group, while its second component $X_2$ by construction is related to the ECI and is associated with the second group (\ed{see also Supplementary Note S4 for further discussion of complexity metric correlations}).

These correlations can be anticipated on theoretical grounds.  The diversity-like quantity $A = \sum_p R_{cp} \pi_p$ we derive is a sum over the activities that a country performs at significant ability levels, weighted by the all-positive elements $\pi_p$.  Similarly, the Fitness metric $F_c = \sum_p M_{cp} Q_p$ is a sum over the activities that a country performs at significant levels with weights given by the all-positive product Qualities $Q_p$.  Not surprisingly, these quantities strongly correlate both with each other and with the count of products in which a country has an RCA greater than 1, $d_c = \sum_p M_{cp}$, a standard measure of diversity.

Similarly, the $b(t)$ coordinate has a close theoretical connection to the ECI and its counterpart metric, the PCI.  The PCI and ECI were proposed by Hidalgo and Hausmann (2009) \cite{Hidalgo2009}, and can be computed with an eigenvector computation as noted by Caldarelli et al. (2012)\cite{Caldarelli2012,Cristelli2013}.  When proximities are measured as $\Phi_{pp'}^{P} \equiv \sum_c \frac{M_{cp} M_{cp'}}{D_c}$, the second vector $\vec{v}_2$ of the dynamical model and the vector of PCIs ($\vec{PCI}$) solve the same eigenvector equation (\ed{see Methods section, ``The eigenvector $\vec{v}_2$ (PCI*)''}), and are therefore identical up to a normalizing constant.  There is ambiguity about which proximity measure one should use to construct the network of activities, yet we find that the second eigenvector is not sensitive to this choice (\ed{see Methods, Fig. \ref{fig_PhiG_correlations}}) and has very high correlation with the vector $\vec{PCI}$ in general.  To distinguish the conventional PCI vector and the second eigenvector of the dynamical model, we will call the latter $\vec{PCI}^* \equiv \vec{v}_2$.  In general, $\vec{PCI}^*$ and $\vec{PCI}$ are strongly correlated, and in the special case $\Phi = \Phi^P$ they are identical.

\begin{figure}[t]
\center
\includegraphics[width=0.8\textwidth]{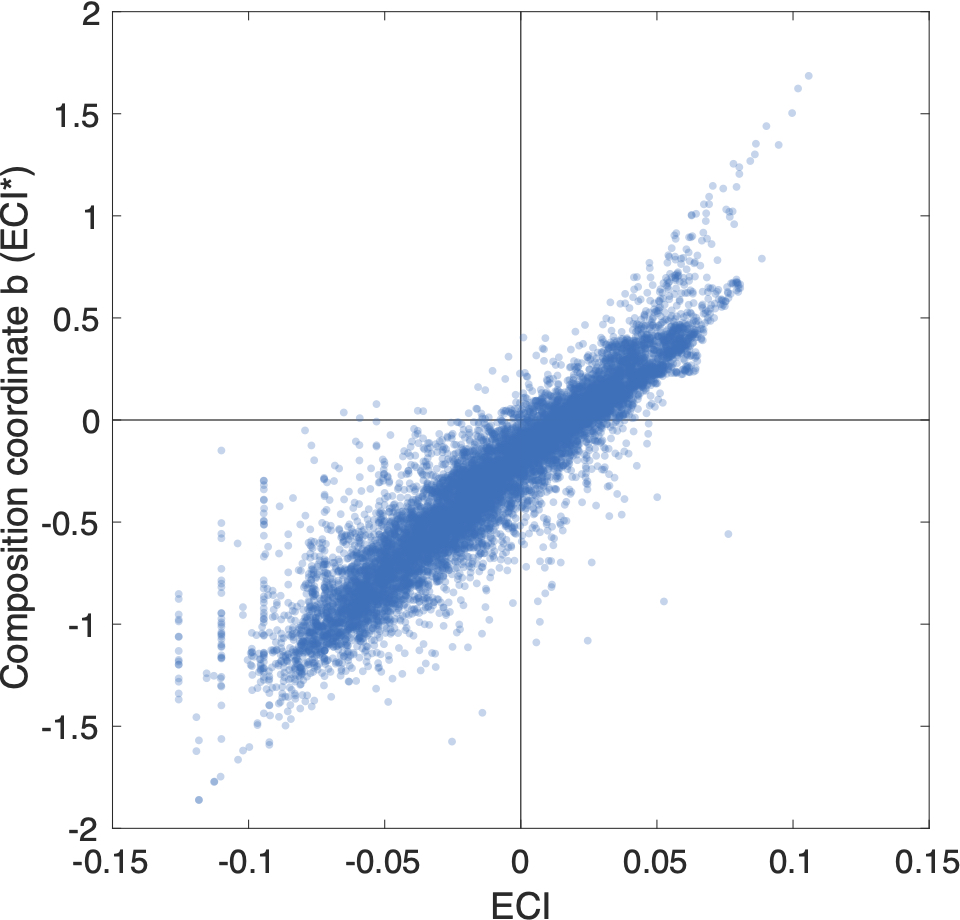}
\caption{The $b$ coordinate (ECI*) versus ECI for 249 countries and regions and 57 years computed from Comtrade data \cite{Comtrade}.}
\label{fig_ECI_v_2ndProjection}
\end{figure}

\begin{figure*}[t]
\center
\sidesubfloat[]{\includegraphics[height=0.175\textheight]{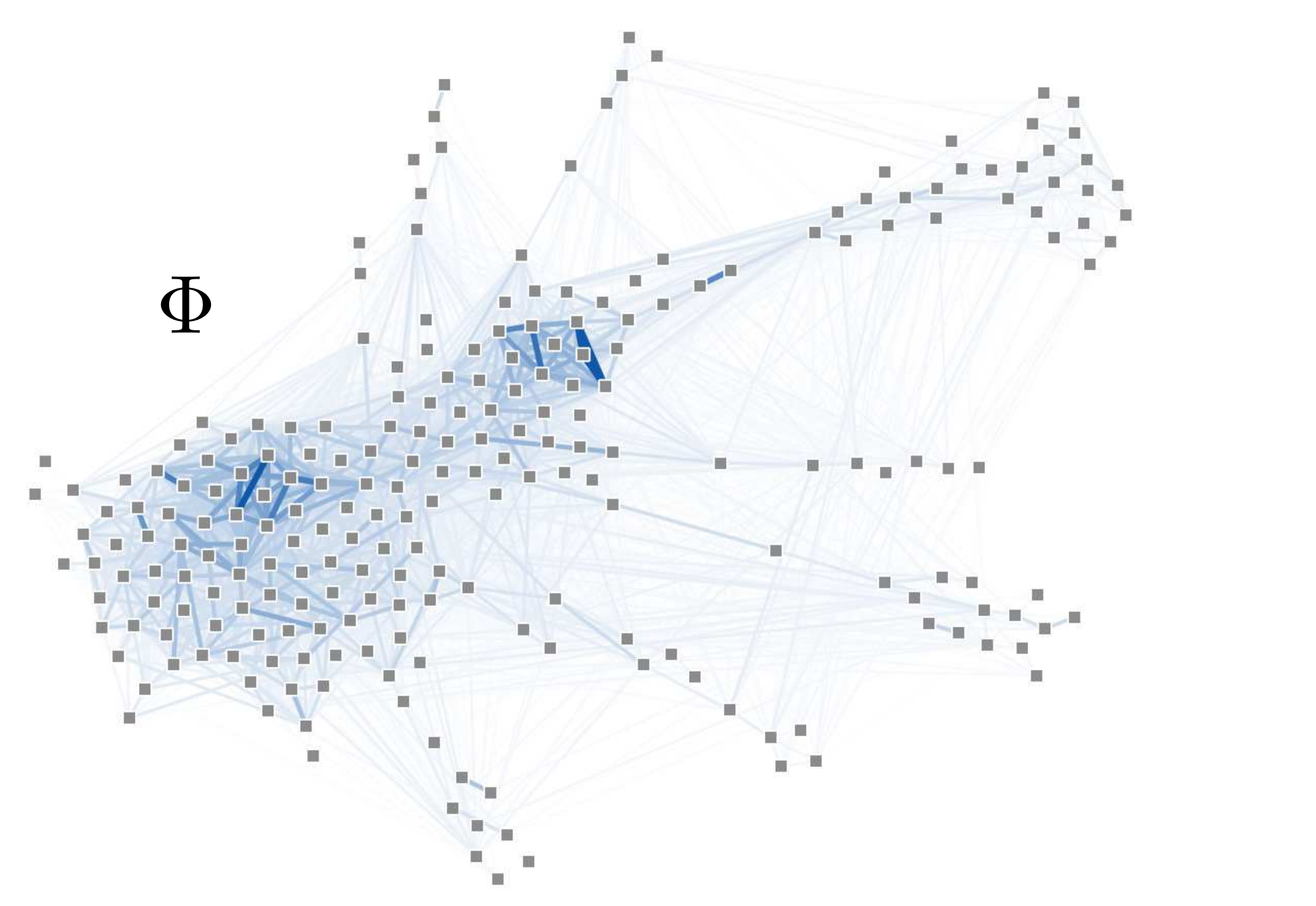}}
\sidesubfloat[]{\includegraphics[height=0.175\textheight]{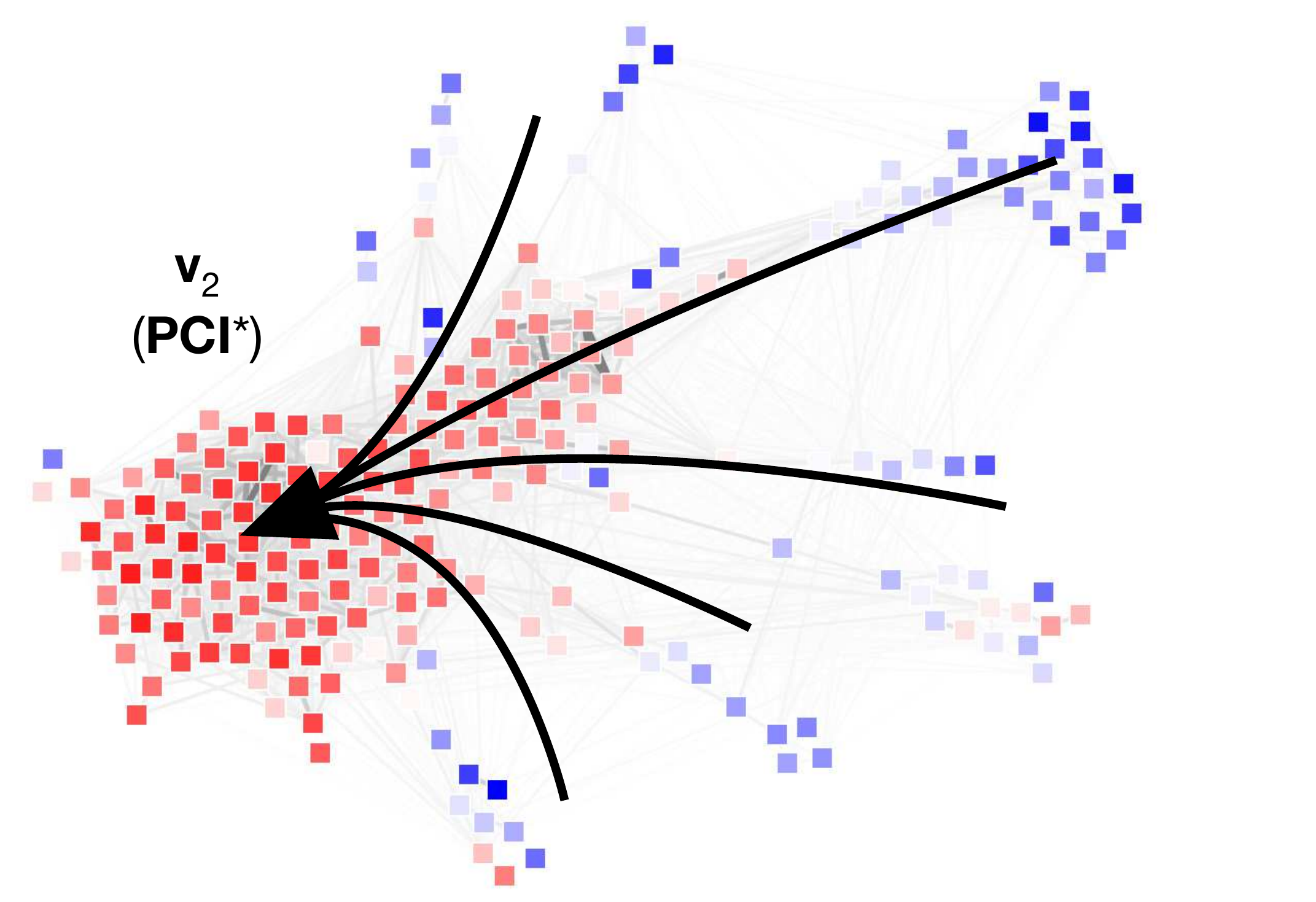}}
\sidesubfloat[]{\includegraphics[height=0.175\textheight]{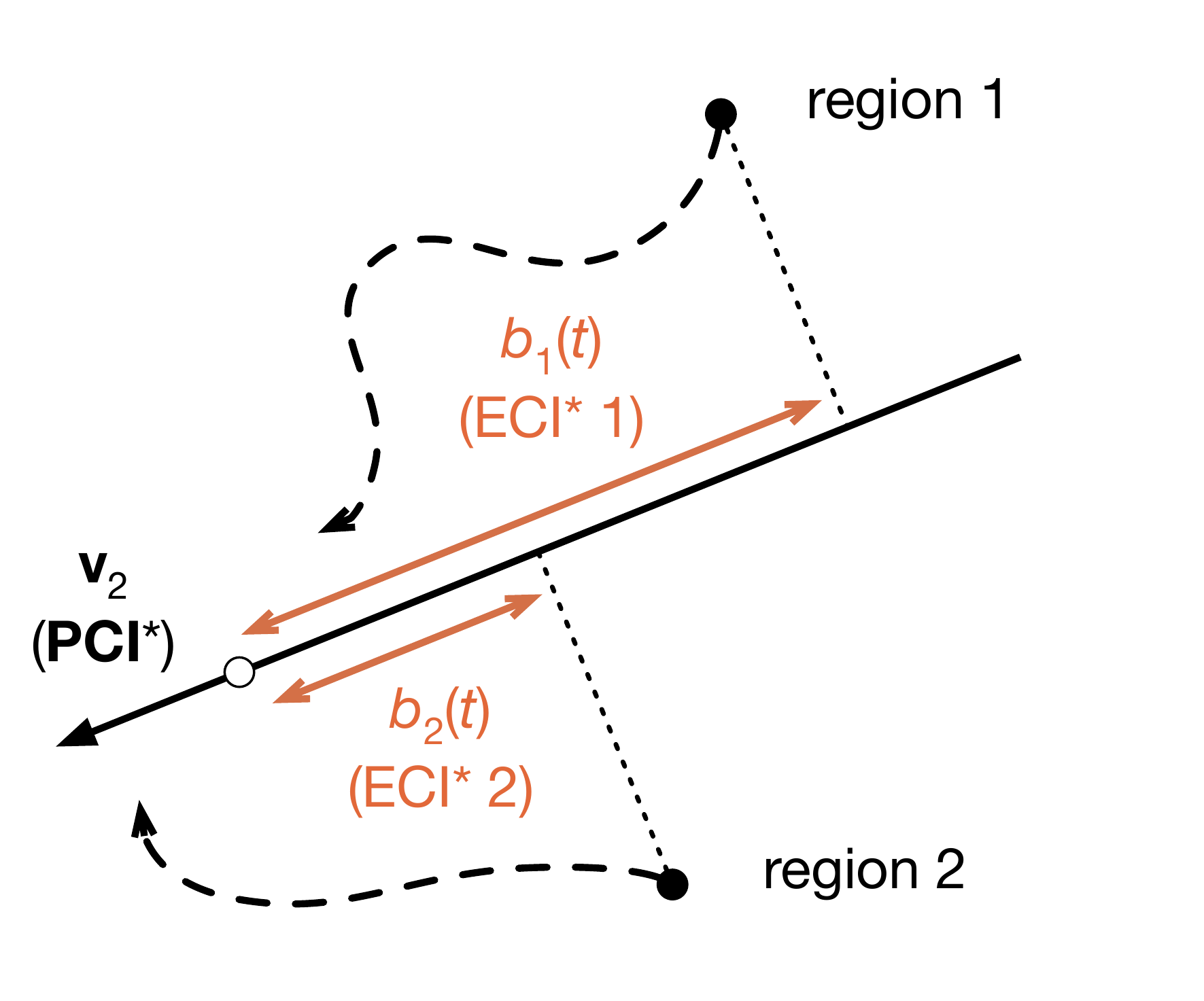}}
\caption{The order in which key quantities are computed in our framework.  \textbf{a} A relatedness network is first defined, on which various patterns of transitions can be described, such as \textbf{b} the compositional mode of change described by the elements of $\vec{v}_2$ ($\vec{PCI}^*$).  Negative values (blue) correspond to economic activities that are vacated in relative terms, while positive values (red) correspond to activities \ed{that} are reached.  \textbf{c} A region's coordinate projection onto this mode determines its value of $b$ (ECI*).}
\label{fig_logicalStructure}
\end{figure*}

The coordinate $b(t)$, which tracks where a country lies on the axis of economic change described by the eigenvector $\vec{v}_2 = \vec{PCI}^*$,  is in turn closely related to the ECI.  We define an ECI-like quantity in our framework, $\text{ECI}^*(t) \equiv b(t)$.  To see that this coordinate is closely related to the ECI, note that a country's ECI is equal to the average PCI of the products in which the country has an RCA greater than 1:\cite{Mealy2019,hill1973reciprocal}
\begin{align}
\text{ECI}_c &= \sum_p \left( \frac{M_{cp}}{\sum_{p'} M_{cp'}} \right) \text{PCI}_{p} 
\label{eq_ECI}.
\end{align}
We can compare this with the ECI*, which can be computed by observing that $b(t) = c_2(t) / c_1(t)$ and calculating the eigen-expansion coefficients $c_2$ and $c_1$.  The resulting calculation (\ed{see Methods, ``The $b$ coordinate (ECI*)''}) yields
\begin{align}
\text{ECI}_c^* = \sum_p \left( \frac{R_{cp} \pi_p}{\sum_{p'} R_{cp'} \pi_{p'}} \right) \text{PCI}_p^*.
\label{eq_ECIstar}
\end{align}
The ECI and $b(t)$ are simply different averages of PCIs or PCI*s.  The conventional ECI gives equal weight to activities in which the country has an RCA greater than 1, and zero weight to other activities.  The ECI* weights activities by their RCA, adjusted by the elements of the first left eigenvector $\vec{\pi} = \vec{w}_1$.  Not surprisingly, these averages are strongly correlated \ed{(Fig. \ref{fig_ECI_v_2ndProjection}), and this holds} for any proximity matrices we consider (\ed{see Methods, Fig. \ref{fig_PhiG_correlations}}).

\paragraph{Interpretation.}
What do we make of these close numerical and theoretical connections between complexity metrics and the PoR-derived structural change coordinates?  Table \ref{tab_comparisonFrameworks} summarizes differences between the usual view of complexity metrics and the dynamical systems view \ed{we present} here based on the PoR.  First, these connections demonstrate (echoing and expanding on Ref. \cite{Mealy2019}) that quantities very similar to complexity metrics can be motivated \ed{through arguments} that have little per se to do with `complexity'.  The relevance of $A(t)$ and $b(t)$ is not tied to considerations of how to infer complexity from data, but to how aptly they characterize long-lived patterns of change in economies.

Second, our results suggest that debates between the main contending metrics could be pointless.  Complexity metrics are typically viewed as being in competition with one another, since they each represent alternative methods to estimate the same underlying quantity (complexity).  But clearly we would not view the coordinates $A(t)$ and $b(t)$ this way; these coordinates just summarize different, complementary information about economic activity baskets, focusing either on their diversity ($A(t)$) or composition ($b(t)$).  The results here affirm the empirical relevance of complexity metrics, while raising the question whether they infer complexity, or essentially summarize major changes in an economy's basket of activities that are associated with development, effectively becoming principled ways to recapitulate and quantify classic statements of structural change.  In the latter case, different metrics emphasize different aspects of development, related either to changes in the diversity or composition of activities.  But as Fig. \ref{fig_phasePlane} illustrated, and classic and recent literature supports \cite{clark1940conditions,Kuznets1957,Imbs2003,Brummitt2020}, both types of changes are general features of economic development.

\begin{table*}[t]
\center
\caption{Comparing the usual view of complexity metrics and the dynamical systems view here based on the Principle of Relatedness.}
\label{tab_comparisonFrameworks}
\resizebox{1\textwidth}{!}{
\renewcommand{\arraystretch}{1.5}
\setlength{\tabcolsep}{5pt}
\begin{tabular}{>{\bf}p{1.5in}p{3.3in}p{3.3in}}
\hline
&	\textbf{Complexity inference view}	
&	\textbf{Dynamical systems / PoR view}\\
\hline					
Goal	
&	Solve an inference problem: Estimate country and product complexities.	
&	Characterize activity baskets: Describe baskets of economic activities in low-order terms using particular coordinates.	\\
\mbox{Competitors or} complements?	
&	Different metrics are in competition.  They offer different methods for solving the inference problem above.	
&	Metrics belonging to different classes in Fig. \ref{fig_complexityMetricCorrelations} are largely complementary.  They summarize different information (diversity, composition) about an economy.	\\
Derivation	
&	Motivated by a data challenge: Infer complexity from data about what goods are produced where.
&	Derived from a model:  Coordinates emerge from a dynamical model of economic diversification.	\\
%
%
\mbox{Role for Principle} of Relatedness?	
&	None in particular.	
&	Yes - the PoR implies the relevance of particular \mbox{coordinates}.	\\
\mbox{Product and} \mbox{country metrics...}	
&	...have similar interpretations: Both measure complexity.  In some setups these metrics solve a set of simultaneous equations.	
&	...have different interpretations: Product metrics collectively describe a direction of economic change, country metrics individually show where a country lies along this axis.  Product metrics are computed first, and country metrics follow.	\\
\hline					
\end{tabular}
}
\end{table*}

Third, the mathematical frameworks used by some complexity metrics could be unnecessary.  Several approaches adopt a framework in which country and product complexity metrics are co-determined by a system of equations \cite{Hidalgo2009,Tacchella2012,Sciarra2020}, i.e.
\begin{align*}
\text{country complexity} &= F(\text{product complexities})\\
\text{product complexity} &= G(\text{country complexities}).
\end{align*}
\ed{This approach is seen as a way to constrain the metrics and give them reasonable properties, with different metrics arising from different choices for functions $F$ and $G$. For example, one obtains the ECI by taking $F$ to be the arithmetic average complexity of the products that a place exports competitively (i.e. with an RCA greater than 1), and $G$ to be the arithmetic average complexity of the countries that competitively export a given product.  Alternatively, one obtains Fitness by replacing $G$ with a harmonic average.} 
But the structure of the \ed{framework} here differs \ed{fundamentally}.  There is no co-determination but a two-step sequence (Fig. \ref{fig_logicalStructure}a-c): According to the PoR, there is a diversification process shaped by a relatedness network, whose eigenvectors capture different directions of change across activities, onto which any given region's coordinate projections can be computed.  This difference in frameworks is related to another -- several product complexity metrics \cite{Hidalgo2009,Sciarra2020} are derived from a similarity matrix between products, while country metrics are derived from a similarity matrix between countries.  But the approach here uses just one of these matrices -- between products.  Under the mindset we adopt to derive our results, this matrix is the fundamental one, capturing technological relationships between products.  The country similarity matrix is incidental, capturing current similarities between countries depending on where they happen to be in their development.

Fourth, our results would help make sense of \ed{conundrums} that so far have been swept under the rug.  For example, oil production is ranked near the bottom of complexity by both the PCI and product Quality measures -- yet oil production is clearly a \ed{complex} activity, requiring knowledge of several fields of engineering, geology, chemistry, physics, transportation and logistics, business operations, and other areas.  At the other end of the spectrum, pottery and works-of-art are ranked highly by these metrics, yet are clearly low-complexity goods.  But these cases are not mysterious if these metrics are not taken as methods to infer complexity, but as measures that track the trajectory of structural change.  Then these rankings would reflect the fact that oil is a difficult product for an economy to diversify away from,%
\footnote{Explanations for the ``resource curse'' have long been debated, see e.g. \cite{ross2015have}.}
%
and that pottery and artwork are activities that are most readily supported in economies that are rich and developed.

Our results invite us to ask whether complexity metrics essentially summarize structural change, rather than act as inference methods that recover hidden information about complexity.  Nevertheless, we cannot rule out the possibility that these metrics play both roles at the same time.  Some recent works explore this by making assumptions about the structure of capabilities and how they are acquired through the process of development \cite{GomezLievano2021,Schetter2019}.  The close correspondence between complexity metrics and long-term change would then reflect the fact that economies become more complex over time, though why and how they do this is not entirely clear. (Presumably, raising productivity or well-being is the goal of development, not complexity per se.)

\paragraph{Dimensionality reduction and related works.}
We are in a position now to discuss several recent works that approach economic complexity from the perspective of dimensionality reduction.  While we emphasized the dynamical meaning of $A(t)$ and $b(t)$, one could also think about these quantities in a dimensionality-reduction framework.

First, note our method of analysis (eigenmode decomposition) is also a dimensionality-reduction technique -- it returns axes (the eigenvectors $\vec{v}_\mu$) in which one can describe high-dimensional data (the matrix of activity baskets $\vec{R}(t)$) along with the coordinate projections of observations onto these axes (coefficients $c_\mu(t)$).  Differences arise in how directions in the vector space are determined and interpreted, as dimensionality-reduction techniques typically aim to identify directions that capture high amounts of variation in data, while eigenmodes characterize coherent patterns in dynamics.  In fact, an eigenmode decomposition not only generates a \emph{representation} of data in a low-dimensional vector space, but shows that the model predicts that countries' activity baskets will \emph{converge} to such a subspace (i.e. corresponding to the directions with the largest time scales).  Differences between countries shrink first among quickly-decaying dimensions, leaving activity baskets scattered primarily along slowly-decaying ones, a well-known outcome of the interaction of structure and dynamics in networks \cite{Simon1961,Schaub2019}.  (See also \ed{Supplementary Note S5} for an expanded discussion.)  

Mealy et al. (2019) \cite{Mealy2019} point out that the conventional ECI can be seen as a dimensionality-reduction tool, equivalent to a spectral clustering algorithm that orders regions along an axis that scores the similarity in their baskets of activities.  Our results are consistent with this view and extend it, showing that the ECI is not just a similarity score but also tracks a classic, long-lived pattern of compositional change.  Our results also \ed{relate to} other, diversity-like complexity metrics, showing how these relate to the dynamics of the PoR.  Sciarra et al. (2020) \cite{Sciarra2020} put forward a complexity index (GENEPY) that reduces the dimensionality of economic activity data in a different way, combining information from two eigenvectors of a country similarity graph.  In one sense, this is related to what we do here, because we also exploit a 2-dimensional picture of country development.  The difference is both technical and conceptual -- the two components that underly GENEPY are taken directly from the elements of the first two eigenvectors of a country-country similarity graph, while the \ed{model on which our framework is based} takes similarities between economic activities to be fundamental, giving rise to eigenvectors that characterize directions of change, and then our coordinates result from projecting countries' export baskets onto these vectors.  Finally, Brummitt et al. (2020) \cite{Brummitt2020} apply machine learning methods to historical export data to extract dimensions that characterize variation in export baskets across countries and time, finding axes that strongly correlate with the PCIs.  This is also closely related to the work here, because the first principal component in Brummitt et al. captures a simultaneous increase in export diversity and a compositional shift, precisely the two kinds of changes that our analytically-derived coordinates separate.

\section{Discussion}\label{sec_discussion}
There are two broad ways to view the results here.  First, our work represents a \ed{dynamical modeling} framework to bridge between short-term and long-term descriptions of economic structural change.  The Principle of Relatedness and its empirical implementations describe structural change on short, year-to-year time scales, and across fine-grained sectors of activities.  In contrast, classic observations of structural change \cite{clark1940conditions,Kuznets1957} emphasize the slow transition of economies over decades across broad economic sectors, along with changes in an economy's diversity \cite{Imbs2003}.  As one would hope, a dynamical framework based on short-term observations is consistent with and bridges into classic observations of long-term change.  In this way our paper elaborates on a long-standing goal of research, dating back at least to Kuznets \cite{Kuznets1957}, to understand the trajectories of economic transformations.

Second, our results tie together two areas of work in the rapidly growing field of economic complexity.  The field's major branches -- studies of relatedness, and complexity metrics -- share concepts and motivating questions, but are linked more in spirit than in math.  The \ed{framework} here connects these branches, suggesting that they refer to short-term and long-term consequences of the same assumptions about economic development.

Our findings highlight that different complexity metrics emphasize different aspects of economic development.  Because of this, our emphasis differs significantly from the bulk of discussions surrounding complexity metrics, which have focused on identifying the correct way to infer `complexity' \cite{Hidalgo2009,Tacchella2012,KempBenedict2014,Mariani2015,Morrison2017,Servedio2018,Teza2018,Schetter2019,Bustos2020,Sciarra2020,ivanova2020measuring,Teza2021,GomezLievano2021} from data.  Our results neither directly support nor contradict the interpretation of complexity metrics as inference methods. Our difference in emphasis partly reflects our strategy, which diverges fundamentally from prior work.  Most work on complexity metrics develops heuristic arguments for using metrics with particular functional forms, but our paper shows how such metrics can be directly derived from an underlying economic model.


Recent work has harnessed new ways of thinking and new analytical methods to study economic development.  Many avenues for further study remain.  \ed{We explored our results in the setting of country exports, but similar analyses could be carried out in networks of occupations \cite{Muneepeerakul2013}, industries \cite{Neffke2011,OClery2019,Hausmann2021}, technology classes \cite{Kogler2013}, research publications \cite{guevara2016research}, or in networks of related locations \cite{Bahar2014},} where the roles of locations and activities could be reversed by projecting the bipartite network of locations and activities onto location nodes instead of activities.  Complexity metrics have been justified in part by their empirical connection to economic growth, though our results suggest these connections may have less to do with complexity per se than with long-term development processes \mbox{more} generally that shape the trajectory of structural change.  All told, our results call for further investigations to improve the dynamical description of structural change  \cite{Coniglio2018}.  We suggest an approach like the one here can be a step toward models that describe these processes with ever-greater fidelity, while being closely tied with empirical metrics, helping shed light on the determinants of growth and development.

\vfil

\section*{Methods}
\edStart
\small
\noindent
Below we first describe theory results that serve as the basis for our analysis, and then describe our data sources and empirical methods.

\paragraph{Relationship between left and right eigenvectors.}\label{eq_stochastic}
Analyzing our dynamical model involves manipulations of the left and right eigenvectors of $L_\Phi$. Here we derive a relationship between these eigenvectors (Eq. \eqref{eq_leftRightEigenvectorRelationship}) that we use to help determine our normalization convention for these vectors (next section), as well as to help derive our ECI-like expression for the $b$ coordinate, Eq. \eqref{eq_ECIstar} (see section ``The $b$-coordinate (ECI*)'' below).

First, note that $L_\Phi = I - \tilde{\Phi}$ and $\tilde{\Phi}$ share the same eigenvectors.  The matrix $\tilde{\Phi} = D^{-1} \Phi$ with $D = \Diag(\Phi \vec{1})$ is a row-normalized stochastic matrix, and so its principal right eigenvector may be taken to be a vector of 1s, $\vec{v}_1 = \vec{1}$.  Its principal left eigenvector can be understood as a vector of stationary probabilities $\vec{\pi}$ for the Markov chain described by $\tilde{\Phi}$, $\vec{w}_1 = \vec{\pi}$.

Define $\Pi = \Diag(\vec{\pi})$.  If $\Phi$ is symmetric, then the left and right eigenvectors of $\tilde{\Phi}$ are related by
\begin{align}
\Pi \vec{v}_\mu = \alpha \vec{w}_\mu
\label{eq_leftRightEigenvectorRelationship}
\end{align}
where $\alpha$ is a constant that depends on the normalization of the eigenvectors.

\textit{Proof:}  By definition the right eigenvectors of $\tilde{\Phi}$ satisfy $\tilde{\Phi} \vec{v}_\mu = \lambda_\mu \vec{v}_\mu$, which can be rearranged as
\begin{align}
(D^{-1/2} \Phi D^{-1/2}) (D^{1/2} \vec{v}_\mu) = \lambda_\mu (D^{1/2} \vec{v}_\mu).\nonumber
\end{align}
The vector $D^{1/2} \vec{v}_\mu$ is thus a right eigenvector of the symmetric matrix $Q \equiv D^{-1/2} \Phi D^{-1/2} = D^{1/2} \tilde{\Phi} D^{-1/2}$.  Similarly, the left eigenvectors of $\tilde{\Phi}$ satisfy $\vec{w}_\mu^\dag \tilde{\Phi} = \lambda_\mu \vec{w}_\mu^\dag$, which can rearranged as
\begin{align}
(\vec{w}_\mu^\dag D^{-1/2}) (D^{-1/2} \Phi D^{-1/2}) = \lambda_\mu (\vec{w}_\mu^\dag D^{-1/2}),\nonumber
\end{align}
showing that $D^{-1/2} \vec{w}_\mu$ is a left eigenvector of the same matrix $Q$.  Finally, since $Q$ is symmetric, any right eigenvector is also a left eigenvector, and thus
\begin{align}
D^{1/2} \vec{v}_\mu = \alpha D^{-1/2} \vec{w}_\mu\nonumber
\end{align}
for some constant $\alpha$.  We therefore have $D \vec{v}_\mu = \alpha \vec{w}_\mu$.  In particular, the first left and right eigenvectors of $\tilde{\Phi}$ are $\vec{v}_1 \propto \vec{1}$ and $\vec{w}_1 \propto \vec{\pi}$, and so we have $D \vec{1} \propto \vec{\pi}$.  Since $D$ is diagonal, in index form this reads $D_{pp} \propto \pi_p$, i.e. $D$ is a matrix that is proportional to $\Pi$, giving us Eq. \eqref{eq_leftRightEigenvectorRelationship}.

\paragraph{Normalization of eigenvectors.}
Our convention for normalizing eigenvectors helps determine the numerical scales of the $A$ and $b$ \ed{coordinates} (and in principle other coordinate projections).  In general left and right eigenvectors are biorthogonal, obeying $W^T V = \Psi$ for some diagonal matrix $\Psi$.  It simplifies many calculations to require that $\Psi = I$ so that left and right eigenvectors obey
\begin{align}
\vec{w}^\dag_\nu \vec{v}_\mu = \delta_{\nu\mu}.
\label{eq_biorthonormality}
\end{align}
This condition does not determine a normalization for the $\vec{w}_\mu$ and $\vec{v}_\mu$ eigenvectors, since these can be rescaled as $\vec{v}_\mu \rightarrow c \vec{v}_\mu$ and $\vec{w}_\mu \rightarrow \frac{1}{c} \vec{w}_\mu$ while preserving Eq. \eqref{eq_biorthonormality}.  However, we can pin down a normalization for the eigenvectors by requiring that they also satisfy Eq. \eqref{eq_leftRightEigenvectorRelationship} with proportionality constant $\alpha = 1$.  One way to accomplish this is to normalize left eigenvectors such that the weighted 2-norm
\begin{align}
||\vec{w} || \equiv \sqrt{\vec{w}^T \Pi^{-1} \vec{w}} \label{eq_Wnorm}
\end{align}
is 1 for each left eigenvector, and to normalize right eigenvectors such that the weighted 2-norm
\begin{align}
||\vec{v} || \equiv \sqrt{\vec{v}^T \Pi \vec{v}} \label{eq_Vnorm}
\end{align}
is 1 for each right eigenvector.  In addition to satisfying Eq. \eqref{eq_leftRightEigenvectorRelationship} with $\alpha = 1$, this normalization convention has the nice side effect that the first right eigenvector $\vec{v}_1 = \vec{1}$ is a vector of 1s, and the elements of the first left eigenvector $\vec{w}_1 = \vec{\pi}$ sum to 1.%
\footnote{We note in passing that Eqs. \eqref{eq_Wnorm} and \eqref{eq_Vnorm} are dual norms.  If $|| \vec{v} ||$ is a norm for vector $\vec{v}$, the dual norm of $\vec{w}$ is the least upper bound of $\vec{w}^T \vec{v}$ for all $\vec{v}$ such that $|| \vec{v} || \leq 1$.  To see that these are dual norms, recall the Cauchy-Schwarz inequality $|\vec{a}^T \vec{b}| \leq || \vec{a} ||_2 || \vec{b} ||_2$.  Making the replacements $\vec{a} = \Pi^{-1/2} \vec{w}$ and $\vec{b} = \Pi^{1/2} \vec{v}$ lets us write this in terms of the norms Eqs. \eqref{eq_Wnorm}-\eqref{eq_Vnorm}:
\begin{align}
|\vec{w}^T \vec{v}| 
\leq || \Pi^{-1/2} \vec{w} ||_2 \,|| \Pi^{1/2} \vec{v} ||_2 
= \sqrt{\vec{w}^T \Pi^{-1} \vec{w}} \; \sqrt{\vec{v}^T \Pi \vec{v}} = || \vec{w} || \; || \vec{v} || .
\end{align}
As $|\vec{w}^T \vec{v}|$  reaches its largest value when $|| \vec{v} || = 1$, the least upper bound of $|\vec{w}^T \vec{v}|$ is $\sqrt{\vec{w}^T \Pi^{-1} \vec{w}}$, which is Eq. \eqref{eq_Wnorm}.}

\paragraph{The $A$ coordinate.}\label{sec_Acoordinate}
The $A(t)$ factor in Eq. \eqref{eq_solution} can be understood as a weighted average of elements of $\vec{R}(t)$.  To see this, first consider the simpler model $\vec{R}(t) = P(t) \vec{R}(0)$ in which $A(t)$ is fixed at 1.  Left-multiplying this model by the first left eigenvector $\vec{w}_1 = \vec{\pi}$, and exploiting the fact that $\vec{\pi}$ is the stationary state of the matrix $P(t) = e^{-L_\Phi (t/\tau)}$, we have
\begin{align}
\bar{R}(t) = \vec{\pi}^T \vec{R}(t) = \vec{\pi}^T \vec{R}(0) = \bar{R}(0).
\end{align}
The left and right sides are weighted averages of elements of $\vec{R}$ with weights given by the elements of $\vec{\pi}$.  This calculation shows that this particular average is unaffected by the multiplication of $\vec{R}(0)$ by $P(t)$, a well-known aspect of consensus dynamics models, which $\vec{R}(t) = P(t) \vec{R}(0)$ is an instance of: The dynamics preserves the average value $\bar{R}(0) = \vec{\pi}^T \vec{R}(0)$ of the initial condition $\vec{R}(0)$ (e.g. \cite{Schaub2019}).

Next, allowing $A(t)$ to vary in Eq. \eqref{eq_solution}, it is clear that the average of the elements of $\vec{R}$ will change by whatever factor $A(t)$ changes.  Left-multiplying Eq. \eqref{eq_solution} by $\vec{\pi}$, after rearrangement we have
\begin{align}
A(t) = \frac{\vec{\pi}^T \vec{R}(t)}{\vec{\pi}^T \vec{R}(0)} = \frac{\bar{R}(t)}{\bar{R}(0)},
\end{align}
showing that $A(t)$ is the factor by which the average of $\vec{R}$ has changed from time $0$ to time $t$.  The time $t = 0$ is just a reference period, with no special significance, and $\bar{R}(0)$ is just a reference value.  Taking this value to be 1 we have simply
\begin{align}
A(t) = \vec{\pi}^T \vec{R}(t) = \bar{R}(t).
\label{eq_Acoordinate_as_average}
\end{align}
The same result can also be obtained from the eigen-expansion Eq. \eqref{eq_eigenmodeDecomposition}.  Left-multiplying by $\vec{w}_1 = \vec{\pi}_1$, and exploiting the biorthogonality of left and right eigenvectors, we have $\vec{\pi}^T \vec{R}(t) = A(t) c_1(0) \vec{\pi}^T \vec{1} = A(t) c_1(0)$, where $\vec{\pi}^T \vec{1} = 1$ because the elements of $\vec{\pi}$ sum to 1.  The $c_1(0)$ coefficient plays a redundant role with $A(t)$; setting it to 1 leaves us with Eq. \eqref{eq_Acoordinate_as_average}.

\paragraph{Review of conventional PCI and ECI}
To aids our discussion of the PCI* and ECI* next we review the calculations that generate the conventional PCI and ECI.  The PCI and the ECI were proposed by Hidalgo and Hausmann (2009) \cite{Hidalgo2009}, and can be computed with an eigenvector computation as noted by Caldarelli et al. (2012)\cite{Caldarelli2012,Cristelli2013}.  Let $n_c$ be the number of countries and $n_p$ the number of products.  Let $R$ be the $n_c \times n_p$ matrix $R$ of RCAs, and let $M$ be a binarized version of this matrix, with elements $M_{cp} = 1$ wherever $R_{cp} \geq 1$ and 0 otherwise.  $M$ may be viewed as an adjacency matrix for a bipartite network connecting countries to products that they export at a significant ability level.  Define the diversity of country $c$'s activity basket as the $c$th row sum of this matrix, denoted as $d_c \equiv \sum_p M_{cp}$. Similarly, define the ubiquity of activity $p$ as the $p$th column sum, denoted as $u_p \equiv \sum_c M_{cp}$.  Let $D$ and $U$ be the diagonal matrices formed from countries' diversities and products' ubiquities, respectively.  One may use the products of $D^{-1} M$ and $U^{-1} M^T$, in different orders, to define two square matrices that eliminate either the country nodes or the product nodes from the bipartite network.  Multiplying these matrices in one order collapses the country dimension, leading to the $n_p \times n_p$ row-normalized stochastic matrix
\begin{align}
\P \equiv U^{-1} (M^T D^{-1} M).
\end{align}
The second right eigenvector of $\P$ defines the unstandardized PCIs, $\vec{PCI}$.  Similarly, collapsing the product dimension leads to the $n_c \times n_c$ row-normalized stochastic matrix
\begin{align}
\C \equiv D^{-1} (MU^{-1} M^T).
\end{align}
The second right eigenvector of $\C$ defines the unstandardized ECIs, $\vec{ECI}$.  Often the PCIs and ECIs are standardized to have zero mean and unit variance across products or countries \cite{Hausmann2014}, but here we work with the unstandardized vectors, and indeed our theory implies that standardizing removes information (see Supplementary Note S6).

\paragraph{The eigenvector $\vec{v}_2$ (PCI*).}\label{sec_eigenvector_two}
The second eigenvector of the Laplacian matrix of the dynamical model, $\vec{v}_2$, is of special interest because it captures the pattern of compositional change with the longest time scale.  Activities that take an especially long time to shift out of have negative elements in this eigenvector, and activities that take an especially long time to reach have positive elements.  This mode of change represents a shift in emphasis from some activities to others, as determined by the signs of elements of the second eigenvector $\vec{v}_2$, thus giving this vector a meaning as a pattern of economic transformation.  We now discuss further how the elements of $\vec{v}_2$ are closely related to the complexity metrics known as the Product Complexity Indices.  To highlight its relation to the PCIs, we refer to this eigenvector as $\vec{PCI}^*$:
\begin{align}
\vec{PCI}^* \equiv \vec{v}_2.
\end{align}

\begin{figure*}[t]
\center
\sidesubfloat[]{\includegraphics[width=0.28\textwidth]{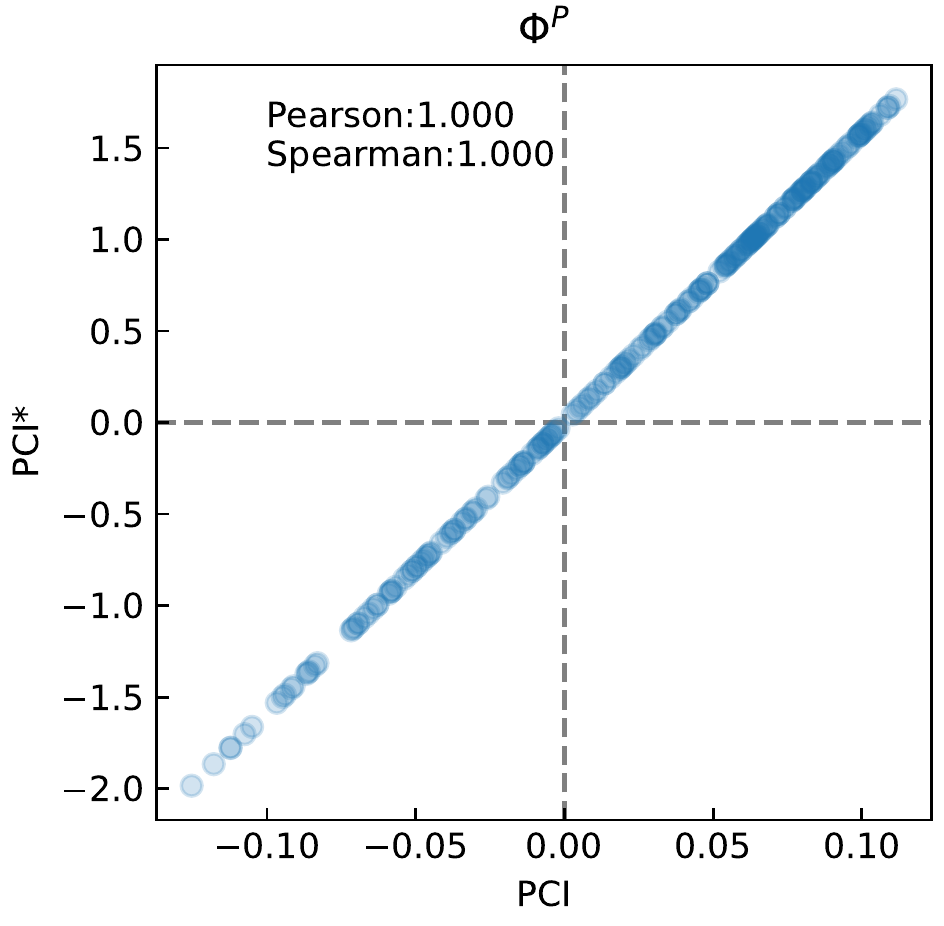}}
\hfil
\sidesubfloat[]{\includegraphics[width=0.28\textwidth]{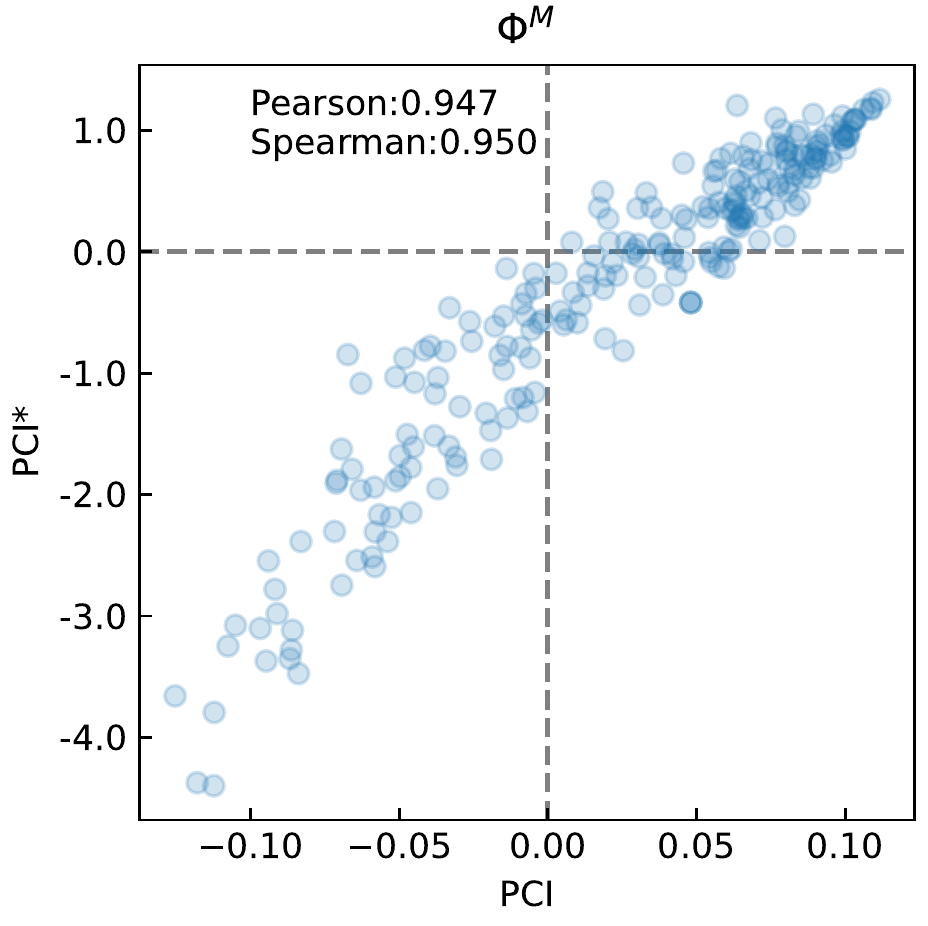}}
\hfil
\sidesubfloat[]{\includegraphics[width=0.28\textwidth]{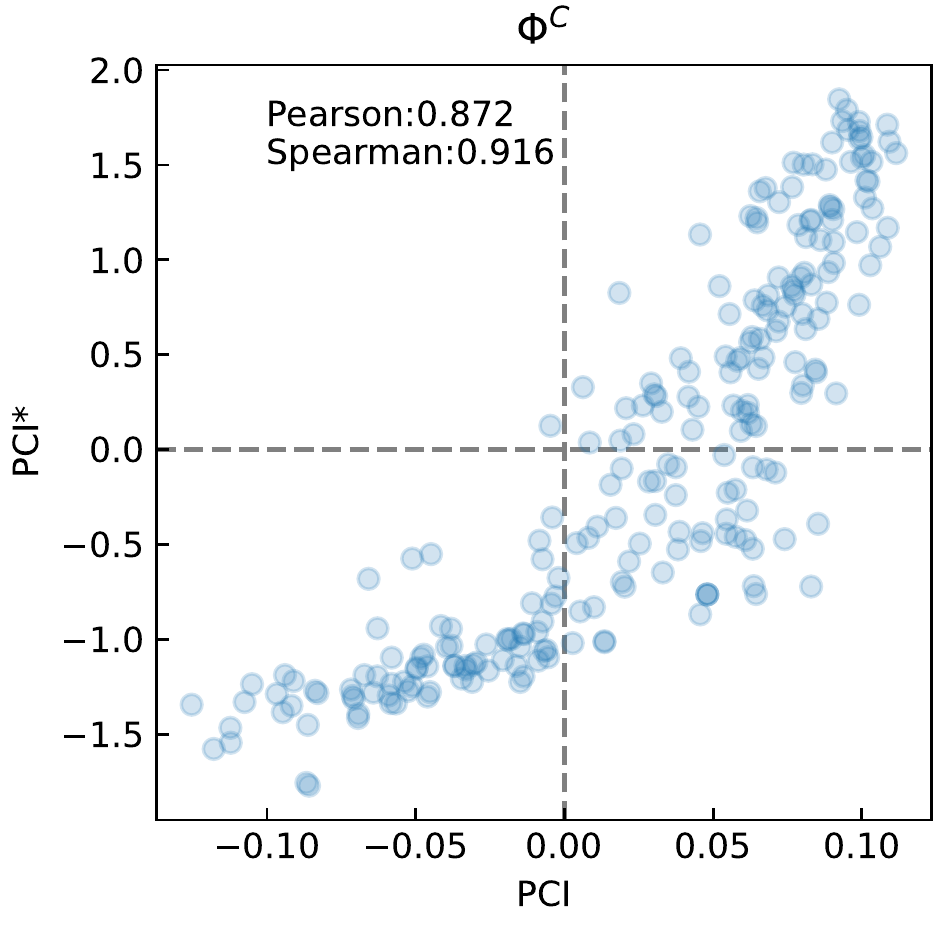}}\\
\sidesubfloat[]{\includegraphics[width=0.28\textwidth]{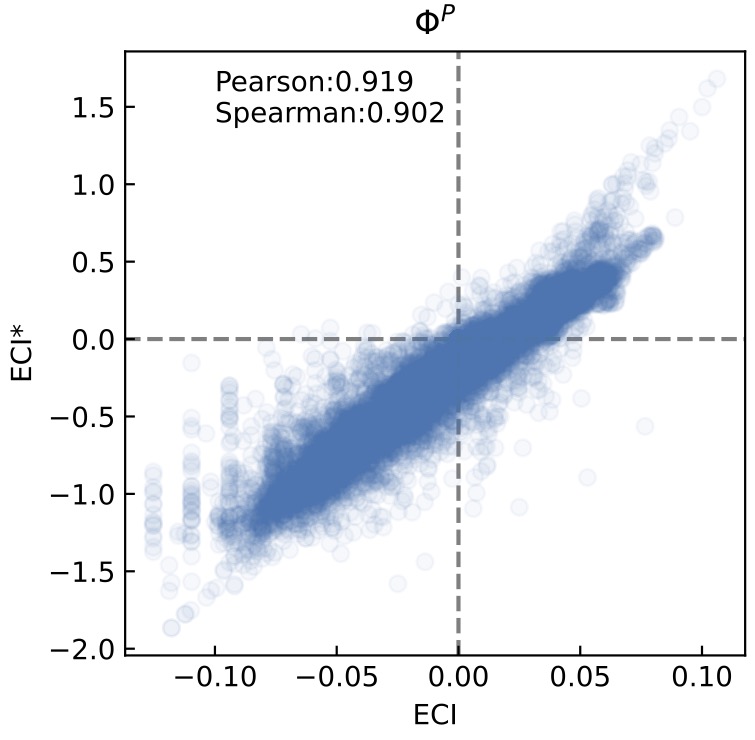}}
\hfil
\sidesubfloat[]{\includegraphics[width=0.28\textwidth]{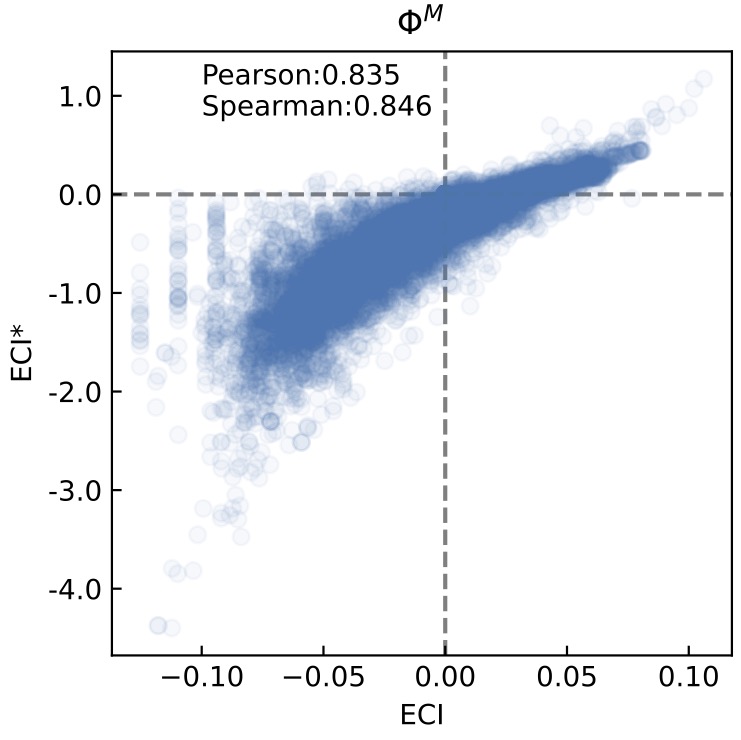}}
\hfil
\sidesubfloat[]{\includegraphics[width=0.28\textwidth]{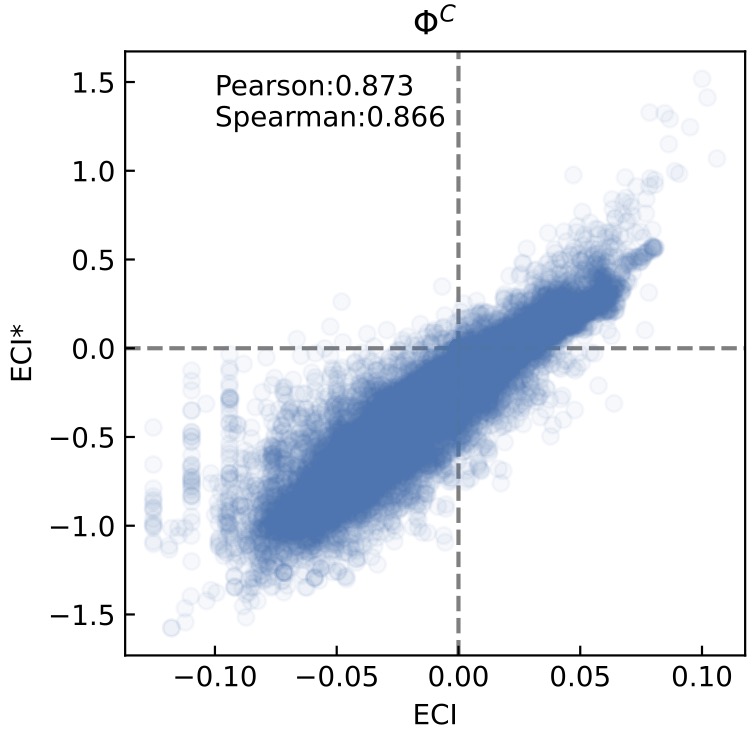}}
\caption{Comparison of PCI* with PCI (\textbf{a-c}) and ECI* with ECI (\textbf{d-f}) for the proximity matrices $\Phi^P$, $\Phi^M$, and $\Phi^C$.}
\label{fig_PhiG_correlations}
\end{figure*}

To see that elements of $\vec{v}_2$ are related to PCIs, first note that the eigenvectors of our dynamical model depend on how we construct the matrix of proximities between activities.  Prior work has generated a variety of proximity matrices, each finding empirical support when used to forecast transitions in economic activities. It is not clear what proximity measure describes the network of transitions best, and our goal here is not to resolve this.  Instead, we show that the structure of the second eigenvector is not sensitive to this choice, and that across a variety of empirically-supported measures of proximity (Fig. \ref{fig_PhiG_correlations}), the second eigenvectors are quite similar to one another, and to the vector of the conventional PCIs.  

First, there is a particular choice of proximity matrix for which $\vec{v}_2 = \vec{PCI}^*$ and the vector of conventional PCIs, $\vec{PCI}$, are identical up to an irrelevant factor.  This happens when the proximities are taken to be 
\begin{align}
\Phi_{pp'}^{P} \equiv \sum_c \frac{M_{cp} M_{cp'}}{D_c},
\label{eq_Phi_PCIimplicit}
\end{align}
or in matrix form $\Phi^P = M^T D^{-1} M$.  If the proximities used in the dynamical model are given by $\Phi = \Phi^P$, then $\vec{PCI}^*$ and $\vec{PCI}$ will be the second eigenvectors of the same eigenvector equation. To see this, note that $\vec{PCI}^*$ is by definition an eigenvector of $L_\Phi = I - \tilde{\Phi}$, and consequently an eigenvector of $\tilde{\Phi}$. Finally, $\tilde{\Phi} \equiv \Diag(\Phi \vec{1})^{-1} \Phi$ equals the matrix $\P$, whose second eigenvector defines the conventional PCIs:
\begin{align}
\tilde{\Phi} 
&\equiv \Diag(\Phi^P \vec{1})^{-1} \Phi^P \nonumber\\
&= \Diag(M^T D^{-1} M \vec{1})^{-1} M^T D^{-1} M \nonumber\\
&= \Diag(M^T D^{-1} \vec{d})^{-1} M^T D^{-1} M \nonumber\\
&= \Diag(M^T \vec{1})^{-1} M^T D^{-1} M \nonumber\\
&= \Diag(\vec{u})^{-1} M^T D^{-1} M \nonumber\\
&= U^{-1} M^T D^{-1} M \nonumber\\
&= \P.
\label{eq_PhiTwiddle_to_scriptP}
\end{align}
Further, a close numerical relation between $\vec{PCI}^*$ and $\vec{PCI}$ carries over beyond this special case; see ``Comparing PCI*/PCI and ECI*/ECI across proximity measures'' and Fig. \ref{fig_PhiG_correlations} below.

\paragraph{The $b$ coordinate (ECI*).}
The $b$ coordinate is associated with the projection of a country's ability vector onto the second eigenvector $\vec{v}_2$.  This coordinate can be written in the form of Eq. \eqref{eq_ECIstar}, showing that it closely resembles the conventional ECI.  To derive this expression, first note that the coefficients of the eigen-expansion can be obtained by exploiting the biorthogonality of the left eigenvectors $\vec{w}_\mu$ with the right eigenvectors $\vec{v}_\mu$. Left-multiplying $\vec{R}(t) = \sum_{\mu} c_{\mu}(t) \vec{v}_\mu$ by $\vec{w}_\mu^\dag$, we have $c_\mu(t) = \vec{w}_\mu^\dag \vec{R}(t)$.  Examining Eq. \eqref{eq_eigenmodeDecomposition}, the first coefficient satisfies (since we have set $c_1(0)$ to 1)
\begin{align}
A(t) = c_1(t) = \vec{w}_1^\dag \vec{R}(t),
\end{align}
and the second coefficient satisfies
\begin{align}
A(t) b(t) = c_2(t) = \vec{w}_2^\dag \vec{R}(t).
\end{align}
It follows that the coordinate $b(t)$ is the ratio of these coefficients:
\begin{align}
b(t) = \frac{c_2(t)}{c_1(t)} = \frac{\vec{w}_2^\dag \vec{R}(t)}{\vec{w}_1^\dag \vec{R}(t)}.
\end{align}

The coordinate $b(t)$ and the activity basket $\vec{R}(t)$ vary across countries.  We now account for this in our notation, which will help us compare $b(t)$ to the ECI.  Let $b_c$ be the coordinate associated with country $c$, and we now expand the vector $\vec{R}$ into the $n_c \times n_p$ matrix of abilities $R = [R_{cp}]$ across countries and products.  In index form, $b_c(t)$ is then
\begin{align}
b_c(t) 
&= \frac{[R(t) \vec{w}_2]_c}{[R(t) \vec{w}_1]_c} \nonumber\\
&= \frac{\sum_p R_{cp}(t) w_{p2}}{\sum_p R_{cp}(t) w_{p1}}.
\end{align}
Next we use three results to express the $b$ coordinate in a different form.  First, in the denominator, we use the fact that the first left eigenvector equals $\vec{\pi}$, $w_{p1} = \pi_p$.  Second, in the numerator, we use the relationship derived earlier between left and right eigenvectors (Eq. \eqref{eq_leftRightEigenvectorRelationship}), $w_{p2} = \pi_p v_{p2}$.  These changes yield
\begin{align}
b_c(t) 
= \frac{\sum_p R_{cp}(t) \pi_p v_{p2}}{\sum_p R_{cp}(t) \pi_p}.
\end{align}
Finally, we insert the definition of PCI*, $\text{PCI}_p^* \equiv v_{p2}$.  The resulting expression has a structure that closely resembles that of Eq. \eqref{eq_ECI}: Taking $\text{ECI}_c^*(t) \equiv b_c(t)$, we have
\begin{align}
\text{ECI}_c^*(t) = \sum_p \left( \frac{R_{cp}(t) \pi_p}{\sum_{p'} R_{cp'}(t) \pi_{p'}} \right) \text{PCI}_p^*.
\end{align}
Like the conventional ECI, Eq. \eqref{eq_ECI}, the ECI* is an average of PCI*s.  The conventional ECI uses uniform weights for all of the nonzero elements in the binarized matrix $M$. The ECI*, in contrast, uses non-uniform weights based on the elements of the unbinarized matrix $R$, and adjusted by the ergodic probabilities $\vec{\pi}$.  \ed{(See SI Note S7 for an interpretation and potential benefits of these weights.)}

We tie the definition of PCI* and ECI* to the Laplacian $L_\Phi$ that governs the dynamics in our diversification model.  This Laplacian could be constructed using different practical measures of proximity, such as $\Phi^{P}$ (Eq. \eqref{eq_Phi_PCIimplicit}), for example, or $\Phi_{pp'}^M \equiv \min\{ (M^TMU^{-1})_{pp'}, (U^{-1}M^T M)_{pp'} \}$, the proximity matrix introduced in Ref. \cite{Hidalgo2007} to define the Product Space.  These different proximity measures represent different guesses for how to \ed{infer} underlying relationships between products, and the propensity for particular product transitions to take place.  Proximity measures yielding better predictions of these transitions could therefore, in principle, produce better calculations of $\vec{PCI}^*$ and ECI* that are more informative of countries' development pathways, offering a \ed{path for refinement of these quantities}.

\paragraph{Description of data.}
We use cleaned UN Comtrade data \cite{Comtrade} publicly available at Harvard Dataverse \cite{Comtrade_Harvard_V4}.  Trade data are reported twice: once as exports by the exporting country and once as imports by the importing country. The data cleaning corrects trade flows to increase the consistency between importer and exporter records of the same flow, and further corrects reported values using an index of reliability based on the consistency of reported values over time.  The data provides total export volumes of products for 249 countries over the period 1962 - 2018 (11,544 export baskets observed across all regions and years).  We analyze these data at the 3-digit SITC product level (235 \ed{product categories}).

\paragraph{Calculation of relatedness network and eigenvectors.}
To obtain our results for country evolution in Fig. \ref{fig_phasePlane} we first compute a matrix of relatedness between products using Eq. \eqref{eq_Phi_PCIimplicit}.  We fix $\Phi$ in the initial year of our data (1962) in accordance with the idea, inherent in the PoR, that economies diversify across a fixed (or at least slow-moving) space of related activities.  From $\Phi$ we compute the row-normalized version of this matrix $\tilde{\Phi}_{pp'} = \Phi_{pp'} / \sum_{p''} \Phi_{p p''}$ as described in the text, whose right ($\vec{v}_\mu$) and left ($\vec{w}_\mu$) eigenvectors are the basis for the remainder of the analysis.  We normalize these eigenvectors with the weighted $L_2$-norms Eqs. \eqref{eq_Wnorm}-\eqref{eq_Vnorm}, though this choice of norms is just a convention and is not consequential, as different choices will simply rescale the coordinates projected onto the eigenvectors.

\begin{figure}[t]
\center
\sidesubfloat[]{\includegraphics[width=0.95\textwidth]{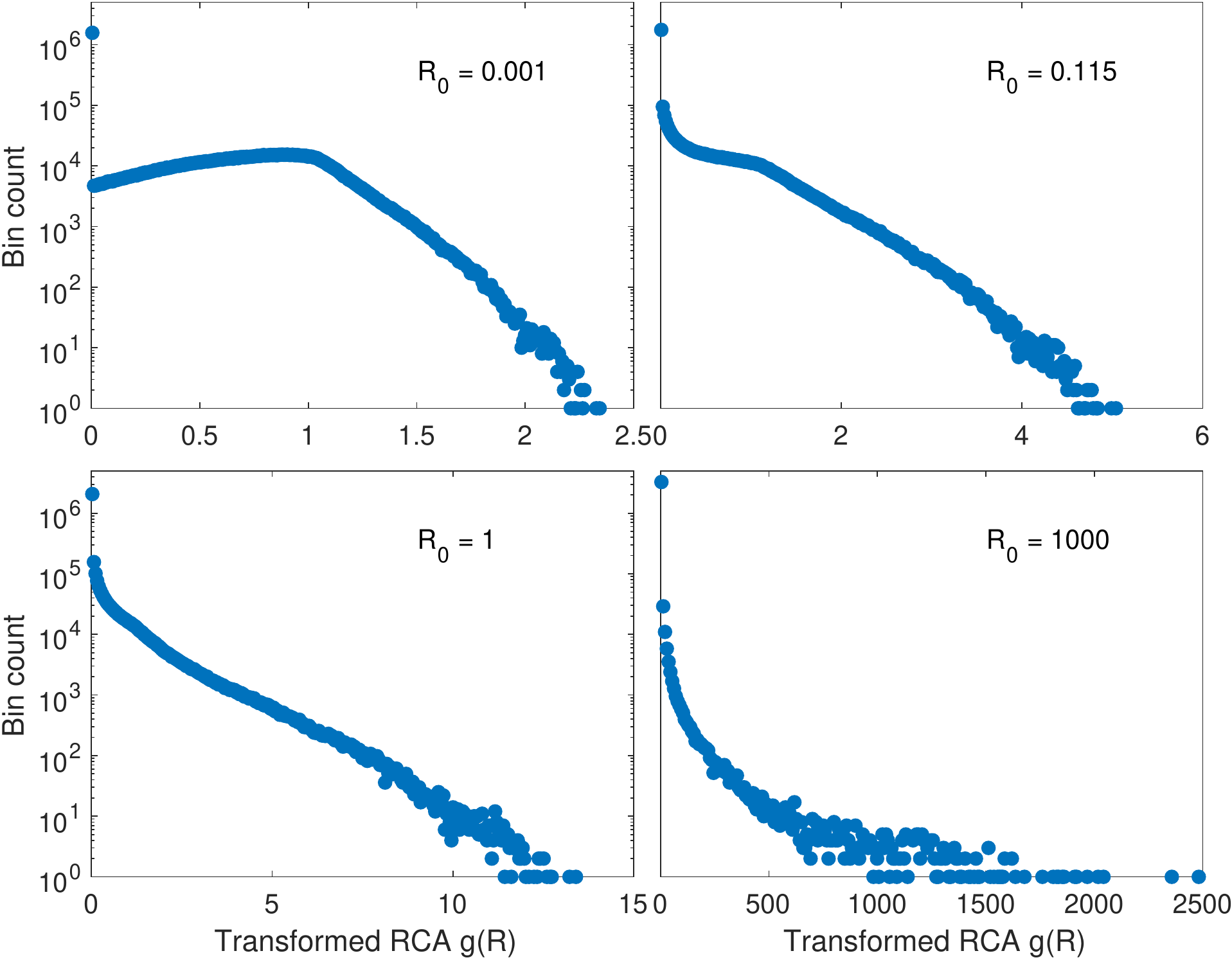}}\\
\vspace{10pt}
\sidesubfloat[]{\includegraphics[width=0.95\textwidth]{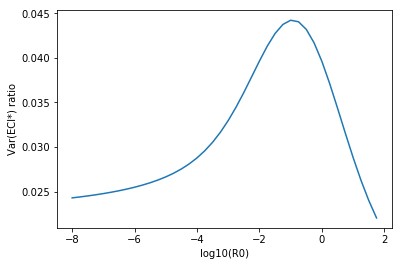}}
\caption{\textbf{a} Distribution of transformed RCAs $g(R')$ for different values of $R_0$.  Small values of $R_0$ weaken the expression of the largest RCAs; large values of $R_0$ do the opposite.  (To make the transformation easier to interpret, we divide the transformation function $g(R')$ by the constant $\log\left(1 + 1/R_0\right)$ so that a raw RCA $R' = 1$ consistently maps to a transformed RCA $R = 1$ for all values of $R_0$.)  \textbf{b} Ratio of variance described by variation in the ECI* (Eq. \eqref{eq_ECIstar_variance}) to total data variability.}
\label{fig_RCAtransformation}
\end{figure}

\paragraph{Calculation of ability vectors $\vec{R}$ and $(A,b)$ coordinates.}
For each country $c$, product $p$, and year we computed the Balassa index of revealed comparative advantage (RCA)\\ $R_{cp}' = (X_{cp} / \sum_{p'} X_{cp'}) / (\sum_{c'} X_{c'p} / \sum_{c',p'} X_{c'p'})$ where $X_{cp}$ is the value of $c$'s exports in product $p$.  RCAs are heavy tailed and are commonly transformed to weaken the effect of extreme values (e.g. \cite{hoen2006measurement,elekes2019foreign,Brummitt2020,Hausmann2021}). Here, we transform them using the function $R = g(R') = \alpha \log(1 + R'/R_0)$.  The constant $\alpha = [\log(1 + 1/R_0)]^{-1}$ is included for convenience so that an RCA of 1 is mapped to 1 on the transformed scale.  This function behaves linearly for small values of $R'$ and logarithmically for large values, and thus achieves the goal of weakening the effect of very large RCAs while also handling RCAs that are identically zero (i.e. which occurs in the many instances where a country has no exports of a product).  The parameter $R_0$ (which sets the transition between the linear and logarithmic regimes) is tuned to maximize the amount of variance in export baskets that the $b$ coordinate explains across time and countries, as we discuss in the next section.  Our results can be reproduced using our obtained value $R_0 = 0.115$.  We then calculate for each country $c$ and time $t$ the RCA vector $\vec{R}_c(t)$ on this transformed scale and use these vectors to compute the $A$ and $b$ coordinates.  The $A$ coordinate for country $c$ and time $t$ is computed as $A_c(t) = \vec{\pi}^T \vec{R}_c(t)$, where $\vec{\pi} = \vec{w}_1$ is the first left eigenvector.  The $b$ coordinate is computed as $b_c(t) = \vec{w}_2^T \vec{R}_c(t) / \vec{w}_1^T \vec{R}_c(t)$, which is equivalent to Eq. \eqref{eq_ECIstar}.

\paragraph{Transformation of RCAs.}
As noted above, we transform raw RCAs with the function $g(R') = \log(1 + R'/R_0) / \log(1 + 1/R_0)$. The parameter $R_0$ sets the transition between the linear and log regimes of the transformation, and modulates the expression of extreme values.  High values of $R_0$ allow large RCAs more expression, while low values of $R_0$ suppress them (Fig. \ref{fig_RCAtransformation}a).  A given value of $R_0$ leads to a given activity vector $\vec{R}(t;R_0)$ and a normalized activity vector $\vec{r}(t;R_0) \equiv \vec{R}(t;R_0) / A(t)$, which has the following representation in the eigenvector basis:
\begin{align}
\vec{r}(t;R_0) = \sum_{\mu} b_{\mu}(t;R_0) \vec{v}_\mu
= \vec{1} + \sum_{\mu \geq 2} b_{\mu}(t;R_0) \vec{v}_\mu.
\label{eq_little_r_eigendecomposition}
\end{align}

We choose $R_0$ to maximize the variance explained across countries and time by the directional vector $\vec{v}_2 = \vec{PCI}^*$, or equivalently, the variance explained by the coordinate $b_2(t;R_0) = b(t) = \text{ECI}^*(t)$.  Our procedure is related to principal component analysis.  The difference is that a PCA \emph{uncovers} a set of variance-maximizing directions in data, while here, the directions are given to us beforehand (the eigenvectors of the Laplacian $L_\Phi$).  Otherwise, we still ask how much variance is explained by data along the particular direction $\vec{v}_2 = \vec{PCI}^*$ for different values of $R_0$ and, like PCA, select a value that maximizes data variability described by this direction.

Let $\vec{r}_i$ be an observation of a normalized activity basket, and let $X = [\vec{r}_i]$ be the $n_p \times N$ data matrix of these observations across countries and years $i \in 1 \ldots N$.  The eigendecomposition Eq. \eqref{eq_little_r_eigendecomposition} corresponds to the matrix factorization
\begin{align}
X(R_0) = V B(R_0),
\end{align}
where $V$ is the $n_p \times n_p$ matrix whose columns are $L_{\Phi}$'s right eigenvectors $\vec{v}_\mu$ and $B$ is the $n_p \times N$ matrix whose $i$th column gives the coordinates of the $i$th observation in the eigenvector basis.  The variances and covariances of the data in the directions of each vector in $V$ can be computed as follows.  Let $X_c = X \Theta$ be the centered (i.e. de-meaned) data where $\Theta \equiv I_N - \frac{1}{N} \vec{1}_N \vec{1}_N^T$ is a centering matrix.  The data covariance matrix is
\begin{align}
C(R_0) &= \frac{1}{N} X_c(R_0) X_c(R_0)^T \nonumber\\
&= \frac{1}{N} X(R_0) \Theta \Theta^T X(R_0)^T \nonumber\\
&= \frac{1}{N} V B(R_0) \Theta \Theta^T B(R_0)^T V^T \nonumber\\
&= \frac{1}{N} V B_c(R_0) B_c(R_0)^T V^T \nonumber\\
&= V \tilde{C}(R_0) V^T,
\label{eq_Cmatrix}
\end{align}
where $B_c = B \Theta$ is the matrix of centered data coordinates in the $V$ basis.  The matrix $\tilde{C} = \frac{1}{N} B_c B_c^T$ is the covariance matrix of the data in the coordinate system given by the eigenvectors of the dynamical model.%
\footnote{As further comparison with PCA, recall that in PCA the data is expressed in a basis $V$ that diagonalizes the covariance matrix.  In such a basis $\tilde{C}$ would be diagonal.  Here, we are expressing the data in a pre-determined basis given by the eigenvectors of $L_\Phi$, and $\tilde{C}$ will have non-zero off-diagonal elements (non-zero covariances in the new coordinates).}

Let $W$ be the $n_p \times n_p$ matrix whose columns are $L_{\Phi}$'s left eigenvectors $\vec{w}_\mu$.  Exploiting the biorthogonality of left and right eigenvectors $W^\dag V = I$, Eq. \eqref{eq_Cmatrix} can be solved for $\tilde{C}$ (the covariances in the $V$ basis) as
\begin{align}
\tilde{C}(R_0) = W^\dag C(R_0) W.
\end{align}
In particular, the variance of activity baskets in the direction $\vec{v}_2 = \vec{PCI}^*$ is
\begin{align}
\text{Var}(\text{ECI}^*;R_0) 
&= \tilde{C}_{22}(R_0) \nonumber\\
&= \vec{w}_2^T C(R_0) \vec{w}_2.
\end{align}
The left eigenvectors were normalized with the modified 2-norm Eq. \eqref{eq_Wnorm}, while principal components are typically normalized with a standard 2-norm.  To remove the scale effect this creates we factor out the standard 2-norm from the left eigenvectors, computing the variance with $\vec{\hat{w}}_2 \equiv \vec{w}_2 / ||\vec{w}_2||_2$:
\begin{align}
\widehat{\text{Var}}(\text{ECI}^*;R_0) &= \vec{\hat{w}}_2^T C(R_0) \vec{\hat{w}}_2.
\label{eq_ECIstar_variance}
\end{align}
For intuition, we could write $\widehat{\text{Var}}(\text{ECI}^*;R_0)$ in terms of the variances explained by principal components.  Let $Z$ be the $n_p \times n_p$ matrix whose columns are principal components.  Inserting $I = ZZ^T$ above leads to
\begin{align}
\widehat{\text{Var}}(\text{ECI}^*;R_0) = \sum_a \left( \vec{\hat{w}}_2^T \vec{z}_a(R_0) \right)^2 \sigma_a^2(R_0).
\end{align}
Thus, the data variance in the direction $\vec{v}_2 = \vec{PCI}^*$ can be written as a weighted sum of the variances of the principal components, with each principal component weighted by its projection ($\vec{\hat{w}}_2^T \vec{z}_a(R_0)$) onto $\vec{v}_2$ in the non-orthogonal basis $V$.  We score the variance that the coordinate $b = \text{ECI}^*$ explains using the ratio of Eq. \eqref{eq_ECIstar_variance} to the total data variability $\sum_a \sigma_a^2(R_0)$.  We calculate this ratio for various values of $R_0$, finding a peak near $R_0 = 0.115$ (Fig \ref{fig_RCAtransformation}b). 

\paragraph{Comparing PCI*/PCI and ECI*/ECI across proximity measures.}\label{sec_comparisonProximities}
To see whether the PCI* and ECI* not only resemble the conventional PCI and ECI theoretically, but also numerically, we compute these quantities with our data and directly compare them.  The PCI* is defined by the second eigenvector $\vec{v}_2$ of the Laplacian matrix $L_\Phi$ and, in general, this and other eigenvectors of our model will vary depending on exactly how the proximity matrix $\Phi$ between activities is constructed.  This matrix has been implemented in a variety of ways that all find empirical support.  Here, we show that the general structure of the second eigenvector is not sensitive to this choice, and that it resembles the vector of conventional PCIs across a range of proximity matrices.  We similarly show that the ECI* numerically resembles the conventional ECI, which is plausible because the ECI* depends on the PCI*.

We consider three proximity matrices.  The first is $\Phi^P$ (Eq. \eqref{eq_Phi_PCIimplicit}).  We also consider the minimum conditional probability-based \cite{Hidalgo2007} proximity measure $\Phi^M$ noted in the main text:
\begin{align}
\Phi_{pp'}^M \equiv \min\{ (M^TMU^{-1})_{pp'}, (U^{-1}M^T M)_{pp'} \}.
\label{eq_PhiM}
\end{align}
We also consider a correlation-based proximity measure \cite{Hausmann2021} $\Phi^C$,
\begin{align}
\Phi_{pp'}^C \equiv \frac{1}{2}(1 + \rho_{pp'}),
\label{eq_PhiC}
\end{align}
where $\rho_{pp'}$ is the Pearson correlation of the RCAs for products $p$ and $p'$ across locations $c$.  Fig. \ref{fig_PhiG_correlations} compares PCI*s to PCIs and ECI*s to ECIs for each proximity matrix.  Pearson and Spearman correlations are shown for each comparison, all with high values between 0.83 and 1.  Panels (a) and (d) show the comparison in the special case $\Phi = \Phi^{P}$ discussed above.  As noted already, when proximities are constructed using Eq. \eqref{eq_Phi_PCIimplicit}, $\vec{PCI}^*$ is equal to $\vec{PCI}$ up to a constant overall factor.  Fig. \ref{fig_PhiG_correlations}d is the comparison of ECI* with ECI in this case, and displays the same results as Fig. \ref{fig_ECI_v_2ndProjection}.  While the $\vec{PCI}^*$ and $\vec{PCI}$ coincide in one special case ($\Phi = \Phi^P$), the same is never true of the ECI* and ECI.  Differences between these quantities will always remain because of the different averaging weights used in Eqs. \eqref{eq_ECI} and \eqref{eq_ECIstar}.  In the closest approach, where $\Phi = \Phi^P$, the Pearson correlation between the ECI* and the ECI is 0.919, and the Spearman rank correlation is 0.902 (Fig. \ref{fig_ECI_v_2ndProjection}).  
In SI Note S6 we discuss why the weights used to calculate ECI* may have certain desirable properties.

Although the closest approach of the $b(t)$ coordinate to the conventional ECI occurs when the proximity matrix is taken to be $\Phi^P$, we note that there is no inherent reason in the dynamical modeling approach presented here to assume this particular proximity matrix.  Proximities between activities could be quantified with other existing matrices, or entirely new ones to be developed, and could be selected based on the quality of forecasts of future transitions in activities, rather than on arguments about how best to infer complexity.

\edEnd

\section*{Acknowledgements}
We thank Ulrich Schetter, R. Maria del Rio Chanona, Ricardo Hausmann, Vito Servedio, François Lafond, \ed{Muhammed Yildirim,} Stefan Thurner, \ed{Doyne Farmer}, \ed{and three anonymous referees} for valuable \ed{feedback}. Frank Neffke acknowledges financial support from the Austrian Research Agency (FFG), project \#873927 (ESSENCSE).

\section*{Data availability}
The datasets analyzed during the current study are available in the Harvard Dataverse, \url{https://dataverse.harvard.edu/dataset.xhtml?persistentId=doi:10.7910/DVN/
H8SFD2&version=4.0}

\section*{Code availability}
The code used for the study is available at \url{https://github.com/complexly/por-structuralchange}.


\end{document}


\title{\bf \large Bridging the short-term and long-term dynamics of economic structural change\\ (Supplementary Information)}
\author{James McNerney, Yang Li, Andr\'{e}s G\'{o}mez-Li\'{e}vano, Frank Neffke}
\date{}
\maketitle

\tableofcontents

\clearpage
\section{Supplementary note: Cancellation of imaginary components in complex conjugate pairs}\label{sec_pairCancellation}
Typically the matrix $\Phi$ \ed{that maps the relatedness of different economic activities is constructed to be symmetric, which results in eigenvalues and eigenvectors of $L_\Phi$ that are real-valued.}  However, a symmetric $\Phi$ is not a requirement of our dynamical theory\ed{;} here we discuss how complex eigenvalues and eigenvectors can be handled without any special difficulties in interpreting the eigenmodes.  The eigenvalues and eigenvectors of real matrices occur in complex conjugate pairs \ed{and real-valued modes can be constructed with a standard trick by combining both components of the pair}.  As noted in the main text, the projection of region $c$ onto the $\mu$th eigenmode is given by $c_\mu(t) = \vec{w}^\dag_\mu \vec{R}(t)$.  If the elements of $\vec{w}_\mu = \vec{a}_\mu + i \vec{b}_\mu$ contain imaginary parts, then it has a paired eigenvector $\vec{w}_{\mu'} = \vec{w}_\mu^* = \vec{a}_\mu - i \vec{b}_\mu$, and  wherever this is the case one may combine the eigenvectors and consider the coefficients in pairs, i.e.
\begin{align}
c_\mu(t) + c_{\mu'}(t) 
= \vec{w}^\dag_\mu \vec{R}(t) + \vec{w}^\dag_{\mu'} \vec{R}(t)
= \left[\vec{w}_\mu^\dag + ( \vec{w}_\mu^\dag )^* \right] \vec{R}(t) 
\end{align}
for which the summed vector $\vec{w}_\mu^\dag + ( \vec{w}_\mu^\dag )^*$ (and therefore the combined projection onto both modes, $c_\mu(t) + c_{\mu'}(t) $) is manifestly real.

\section{Supplementary note: Individual country trajectories}

\begin{figure}[h]
\center
\includegraphics[width=0.95\textwidth]{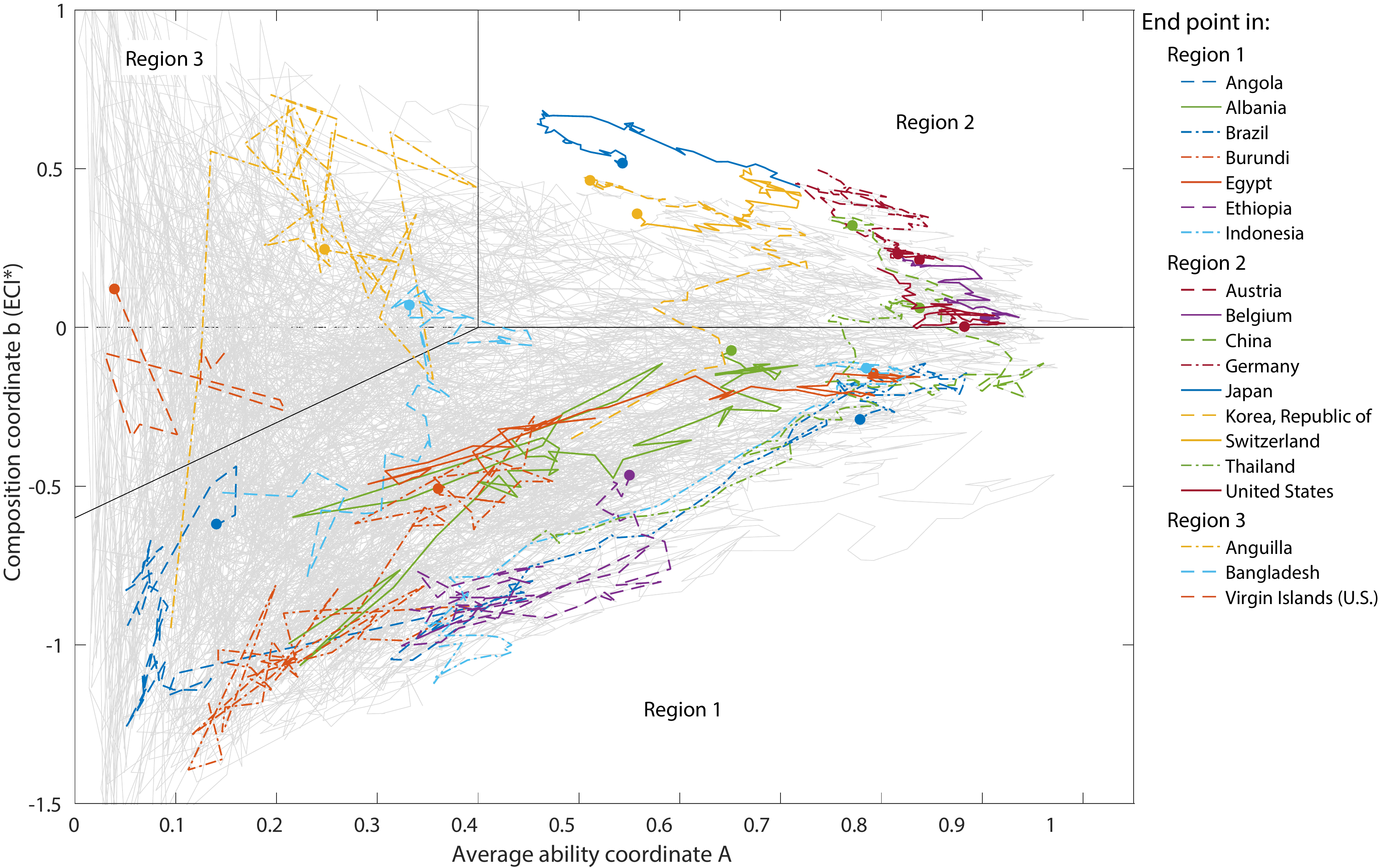}
\caption{Trajectories of selected countries in the $A$-$b$ phase plane.  The labels Region 1, Region 2, and Region 3 refer to the same parts of the plane as in Figure 4e of the main paper. Round markers identify the end point of each trajectory.}
\label{fig_trajectories}
\end{figure}

\ed{Fig. \ref{fig_trajectories} shows the individual trajectories of countries in our data from which the average directional movement depicted in Fig. 3e of the main text was derived.  In some parts of this plane only a few countries have been observed, and in these sparse areas individual countries can dominate the behavior of directional lines in Fig. 3e.  An instance of this small-numbers effect is seen in Japan's movements in Region 2.}

\section{Supplementary note: Higher-order eigenvectors and eigenvalues}
\begin{figure}[h]
\center
\sidesubfloat[]{\includegraphics[width=0.46\textwidth]{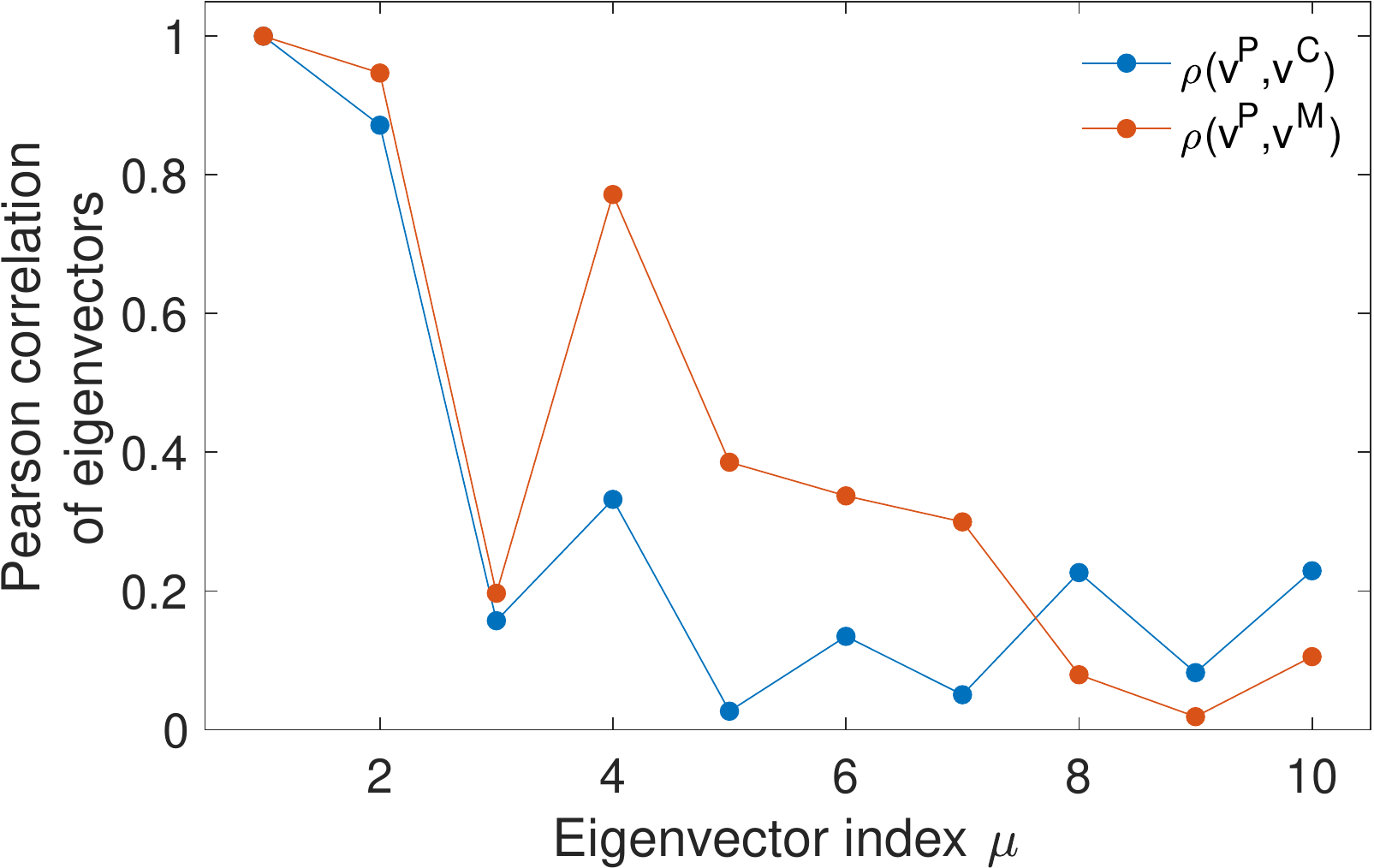}}\\
\vspace{12pt}
\sidesubfloat[]{\includegraphics[width=0.46\textwidth]{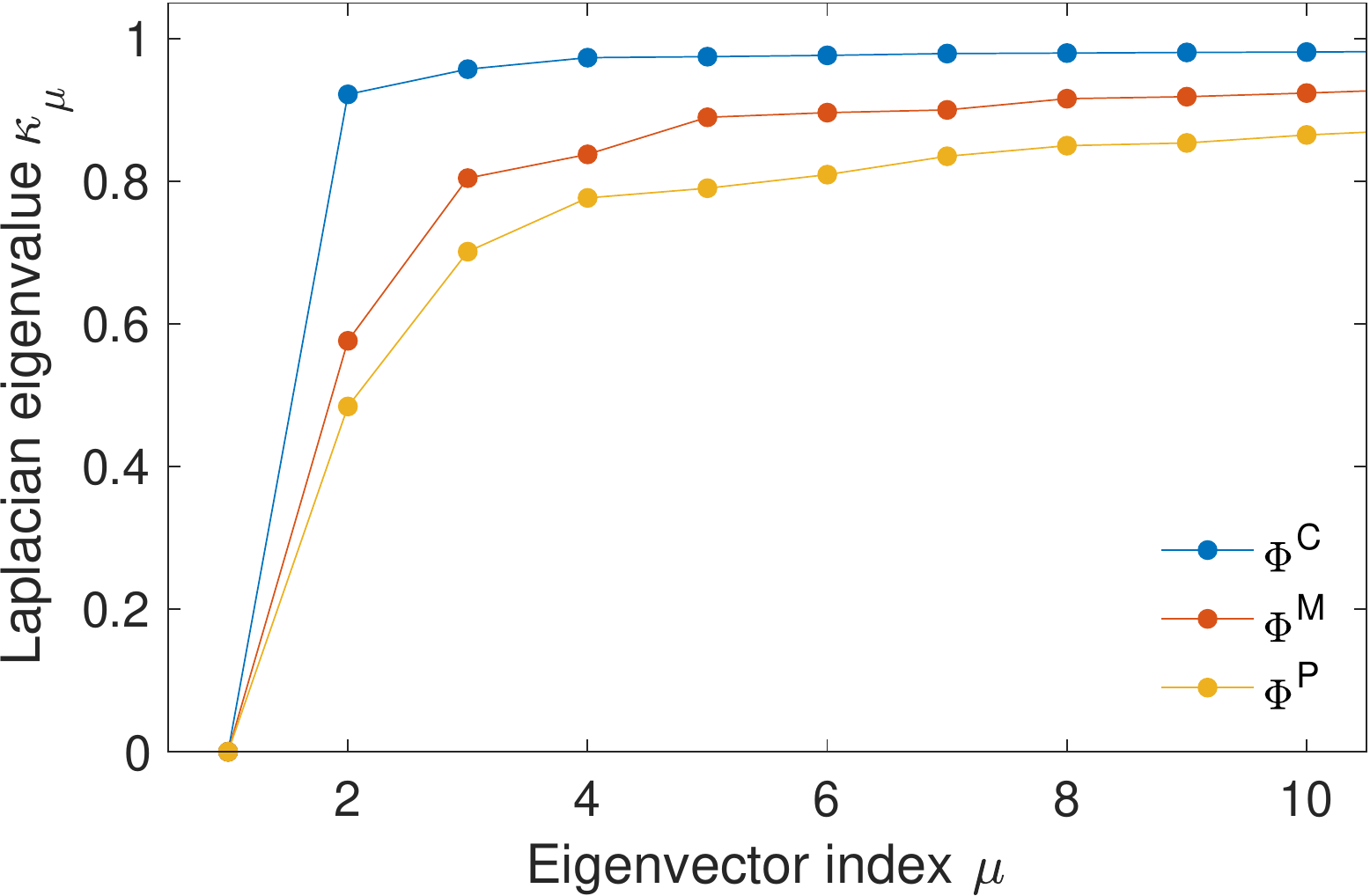}}
\caption{\textbf{a} Pearson correlations showing how eigenvectors of different proximity matrices become less similar to one another as one moves to higher-order eigenvectors. The series labeled $\rho(\vec{v}^P, \vec{v}^C)$ gives correlations between eigenvectors of $\tilde{\Phi}^P$ and $\tilde{\Phi}^C$, and the series labeled $\rho(\vec{v}^P, \vec{v}^M)$ gives correlations betweens eigenvector of $\tilde{\Phi}^P$ and $\tilde{\Phi}^M$. \textbf{b} Variation of the first 10 eigenvalues of the graph Laplacian $L_\Phi$ for the proximity metrics $\Phi^P$, $\Phi^M$, and $\Phi^C$.  }
\label{fig_eigenmodeComparison}
\end{figure}

\edStart
Here we examine the empirical structure of the right eigenvectors and eigenvalues of $\tilde{\Phi}$ for the three choices of proximity matrix $\Phi$ describe in the Methods ($\Phi^P$, $\Phi^M$, and $\Phi^C$). Higher-order eigenvectors (i.e. those beyond the second eigenvalue) are of interest because in principle they could describe other modes of economic transformation besides the particular diversity and compositional changes that we study in the main text.  However, we find that the proximity matrices in the literature only have strong agreement in the structure of the first two eigenvectors, and do not straightforwardly resolve how many eigenvectors matter.  We row-normalize each proximity matrix to obtain the matrices $\tilde{\Phi}^P$, $\tilde{\Phi}^M$, and $\tilde{\Phi}^C$ and examine the eigenvectors of the resulting graph Laplacian $L_\Phi$ (i.e. $L_\Phi = I - \tilde{\Phi}^P$, $L_\Phi = I - \tilde{\Phi}^M$, or $L_\Phi = I - \tilde{\Phi}^C$).  We put their eigenvectors in ascending order by eigenvalue $\kappa_\mu$, and use the eigenvectors of $\tilde{\Phi}^P$ as a reference point, computing the correlation of the $\mu$th eigenvector of $I - \tilde{\Phi}^P$ with the $\mu$th eigenvector of $L_\Phi = I - \tilde{\Phi}^M$ or $L_\Phi = I - \tilde{\Phi}^C$ (Fig. \ref{fig_eigenmodeComparison}a).  The first eigenvectors have a correlation $\rho = 1$ since $\tilde{\Phi}^P$, $\tilde{\Phi}^M$, and $\tilde{\Phi}^C$ share the same first right eigenvector, $\vec{v}_1 = \vec{1}$.  The first two eigenvectors have very high correlations, but the correlations fall off beyond this (with a notable spike for the 4th eigenvector).  We see related issues when we consider the eigenvalues (Fig. \ref{fig_eigenmodeComparison}b).  The spectra derived from different proximity matrices differ not only in their values but also in the number of eigenvalues that appear significantly different from others.

Resolving the structure of higher-order eigenvectors and estimating how many eigenvectors appear relevant for long-term structural transformation means addressing the question of what the right proximity matrix is.  Future work could, for example, investigate which proximity matrices show the highest predictive ability \cite{li2022relatedness}.  Alternatively, one could avoid directly constructing the proximity matrix $\Phi$ and adopt an agnostic approach such as Dynamical Mode Decomposition, which involves estimating the best-fitting linear operator that advances a system forward in time.
\edEnd

\FloatBarrier
\section{Supplementary note: Correlations of coordinates with complexity metrics}
\begin{figure}[h]
\center
\sidesubfloat[]{\includegraphics[width=0.47\textwidth]{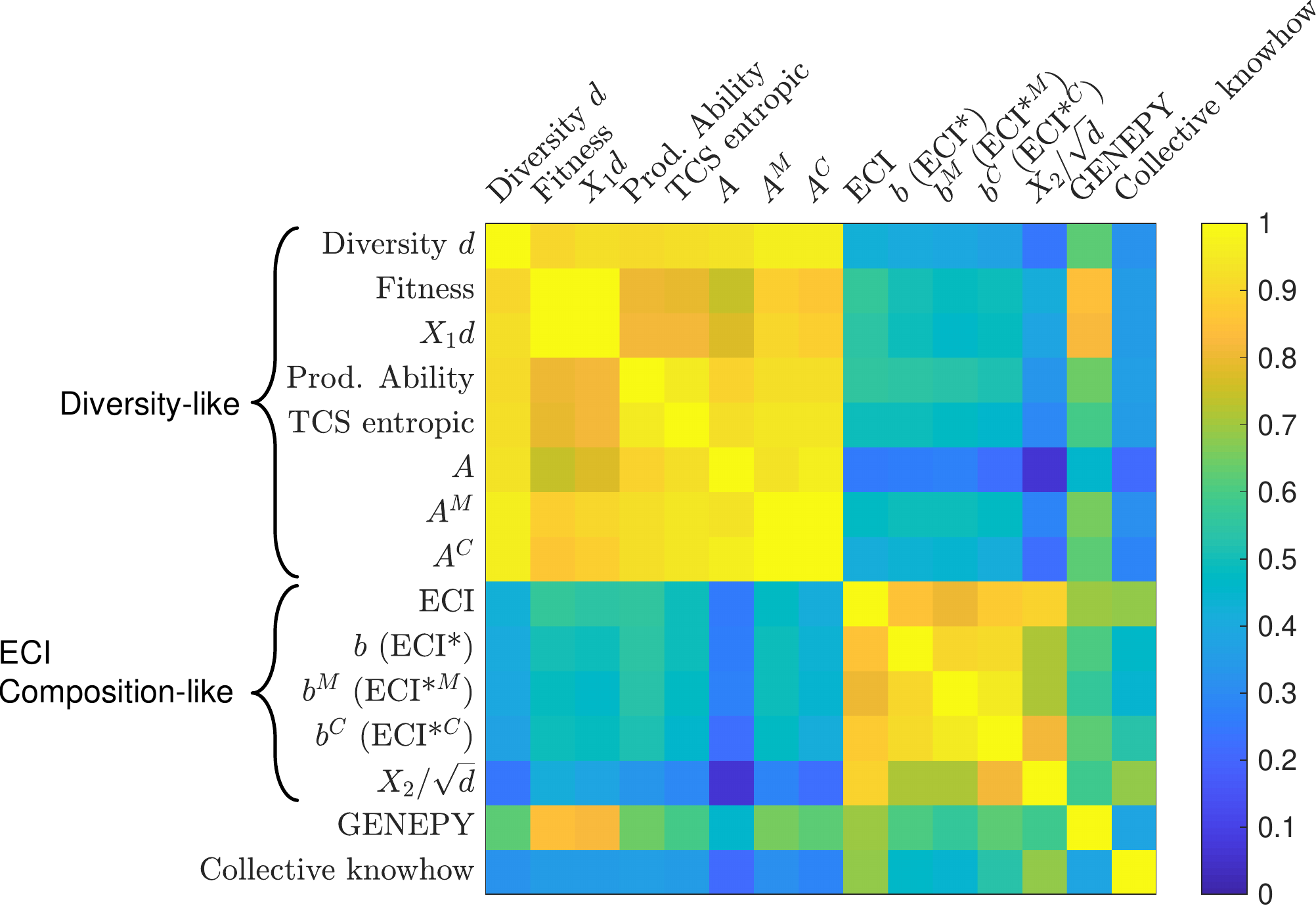}}
\hfill
\sidesubfloat[]{\includegraphics[width=0.47\textwidth]{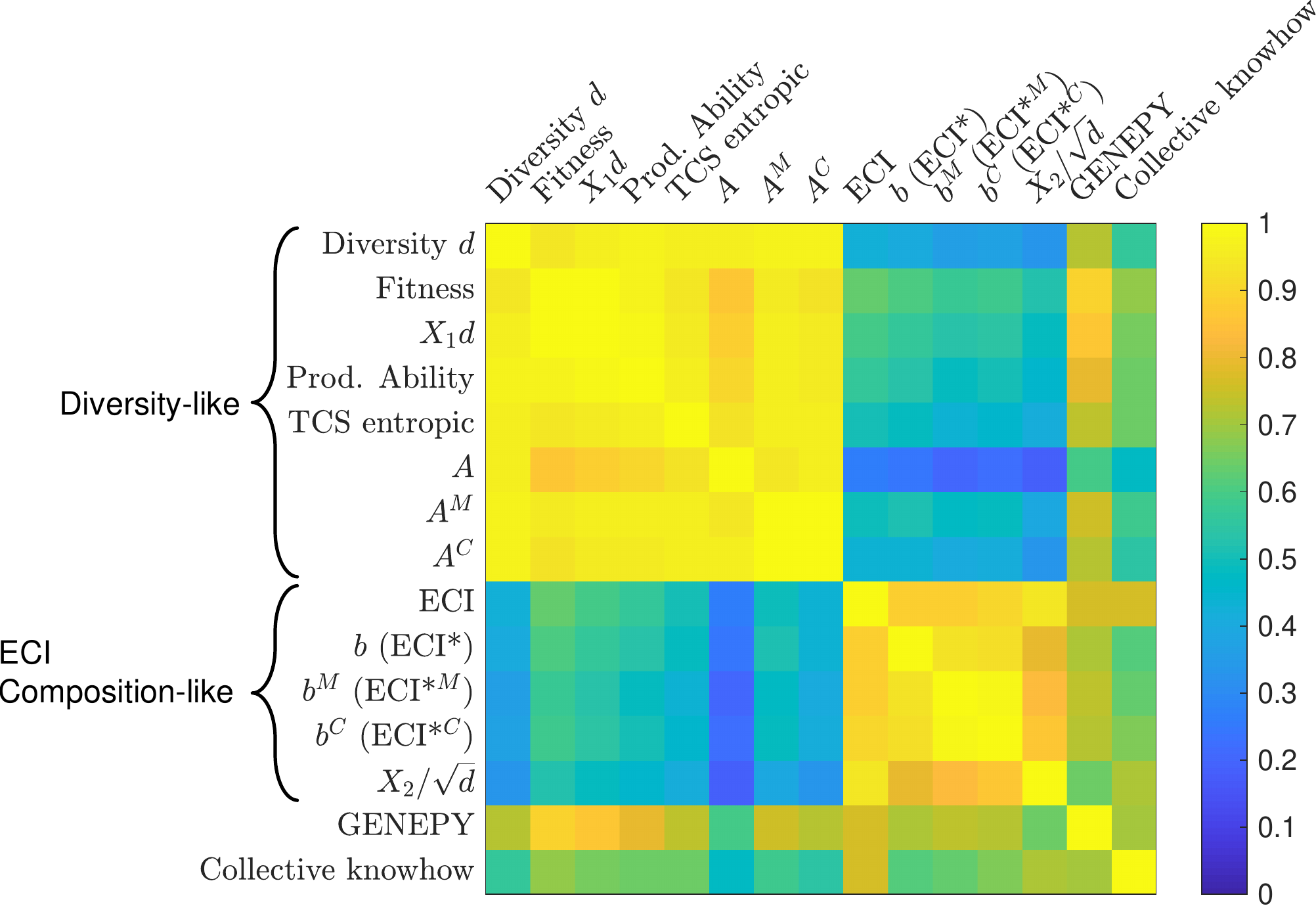}}
\caption{Correlations among our $A$ and $b$ (ECI*) coordinates, several proposed measures of complexity (ECI, Fitness, Production Ability, the entropic measure of Teza, Caraglio, and Stella, and GENEPY), and related quantities. \textbf{a} Pearson correlations.  \textbf{b} Spearman rank correlations (identical to Fig. 5 in the main text).  The coordinate $A$ is our average ability coordinate used throughout the text, defined using the proximity matrix $\Phi^P$, while $A^M$ and $A^C$ refer to the same coordinate obtained from using the proximity matrices \ed{$\Phi^M$ and $\Phi^C$} respectively.  All quantities were computed for the year 2016 using UN Comtrade data.}
\label{fig_complexityMetricCorrelations}
\end{figure}

Here we further discuss connections of the $A$ and $b$ coordinates with complexity metrics.  As emphasized in the main text, nearly all complexity metrics fall into two groups (Fig. \ref{fig_complexityMetricCorrelations}) that correspond to the two coordinates $A$ and $b$ in our dynamical theory.  The \ed{main text discusses} theoretical and empirical similarities of the $A$ coordinate with country Fitness \cite{Tacchella2012} and of the $b$ coordinate with the ECI \cite{Hidalgo2009}.  Two other metrics, Production Ability \cite{Bustos2020} and the entropic measure of Teza, Caraglio, and Stella \cite{Teza2018,Teza2021}, are also strongly correlated with the $A$ coordinate.  A recently introduced hybrid metric known as GENEPY \cite{Sciarra2020} contains aspects of both the ECI and country Fitness.  GENEPY is computed by first obtaining the two eigenvectors with the largest eigenvalues of a country-to-country similarity matrix.  Letting $X_{c,1}$ and $X_{c,2}$ denote the elements of these eigenvectors, and letting $\lambda_1$ and $\lambda_2$ denote the eigenvalues, the GENEPY of country $c$ is $\textit{GENEPY}_c = \left( \sum_{i=1}^2 \lambda_i X_{c,i}^2 \right)^2 + 2 \sum_{i=1}^2 \lambda_i^2 X_{c,i}^2$.  By design, $X_{c,1}$ is closely related to Fitness and $X_{c,2}$ is closely related to the ECI.  Sciarra et al. (2020) show that, for example, when the country-country proximity matrix is $N_{cc'} \equiv \sum_p \frac{M_{cp} M_{c'p}}{d_c d_{c'} (k_p')^2}$ where $k_p' = \sum_c M_{cp} / k_c$, then $X_{c,1} d_c = F_c$, and $X_{c,2} / \sqrt{d_c}$ is strongly correlated with $\text{ECI}_c$.  We find empirically that the $X_{c,1}$ and $X_{c,2}$ components underlying GENEPY also have similar relationships to the $A$ and $b$ coordinates, i.e. $X_{c,1} d_c$ is highly correlated to $A$, and $X_{c,2} / \sqrt{d_c}$ is highly correlated to $\text{ECI}^*$.  Of the metrics we examine, only collective knowhow \cite{GomezLievano2021} shows no similarly strong empirical relationship to the $A$ and $b$ coordinates.  The existence of these two groups emphasizes that, when \ed{viewed through the analytical lens of our dynamical modeling framework}, \ed{most} existing complexity metrics emphasize different aspects of the same development process: either changes in economic diversity, or a particular pattern of compositional shifts in activities.  However, as Fig. 2 in the main paper illustrated, both aspects of change are important characteristics of economic development.

Further comments are in order regarding the GENEPY metric.  Sciarra et al. highlight that GENEPY merges information about countries from coordinates on two axes.  Interestingly, we also arrived at a representation of country development on a 2D plane, using a very different starting point for our calculations.  GENEPY is motivated by the goal of reconciling different methods for inferring complexity from data within a consistent linear algebra and networks framework.  In contrast, our 2D plane arose from our analysis of a dynamical model based on the Principle of Relatedness.  We note three differences in the resulting planes that define each of our approaches.  First, as noted already above, our $A$ and $b$ coordinates correspond most closely to $X_{c,1} d_c$ and $X_{c,2} / \sqrt{d_c}$ rather than Sciarra et al.'s $X_{c,1}$ and $X_{c,2}$ themselves.  Second, a fundamental input in the method of Sciarra et al. is a country-country proximity matrix $N_{cc'}$.  The elements of the eigenvectors of this matrix then directly give countries' coordinates in the plane.  In contrast, the fundamental input in the framework here is a product-product proximity matrix $\Phi_{pp'}$.  Our approach comes with a two-step process: Eigenvectors derived from $\Phi$ first define directions of movement, and then country coordinates are obtained from projections onto these axes.  Third, Sciarra et al. interpret their coordinates in terms of their relationship to Fitness and the ECI.  Here, we exploit the fact that, by analyzing the changes in the world economy captured by movement in the directions $\vec{v}_1$ and $\vec{v}_2$, we can give economic meaning to the coordinates $A$ and $b$ (ECI*) that define our 2D plane, as capturing changes in average ability or shifts in activity composition.

\begin{figure}[t]
\center
\sidesubfloat[]{\includegraphics[width=\textwidth]{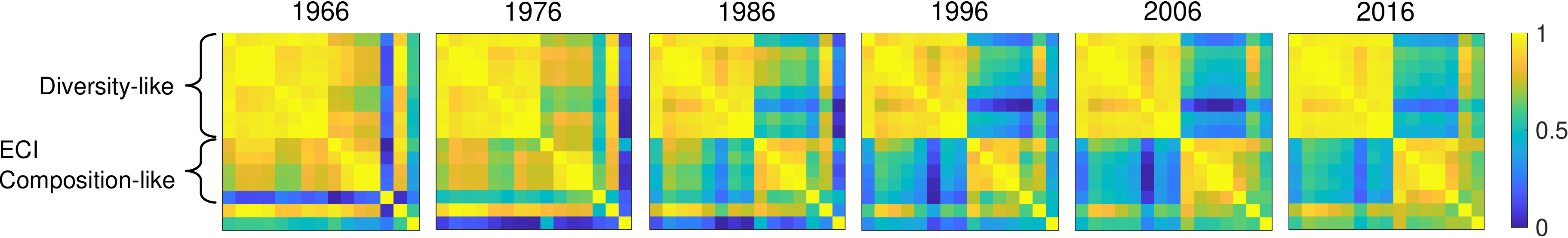}}\\
\vspace{10pt}
\sidesubfloat[]{
\includegraphics[width=0.35\textwidth]{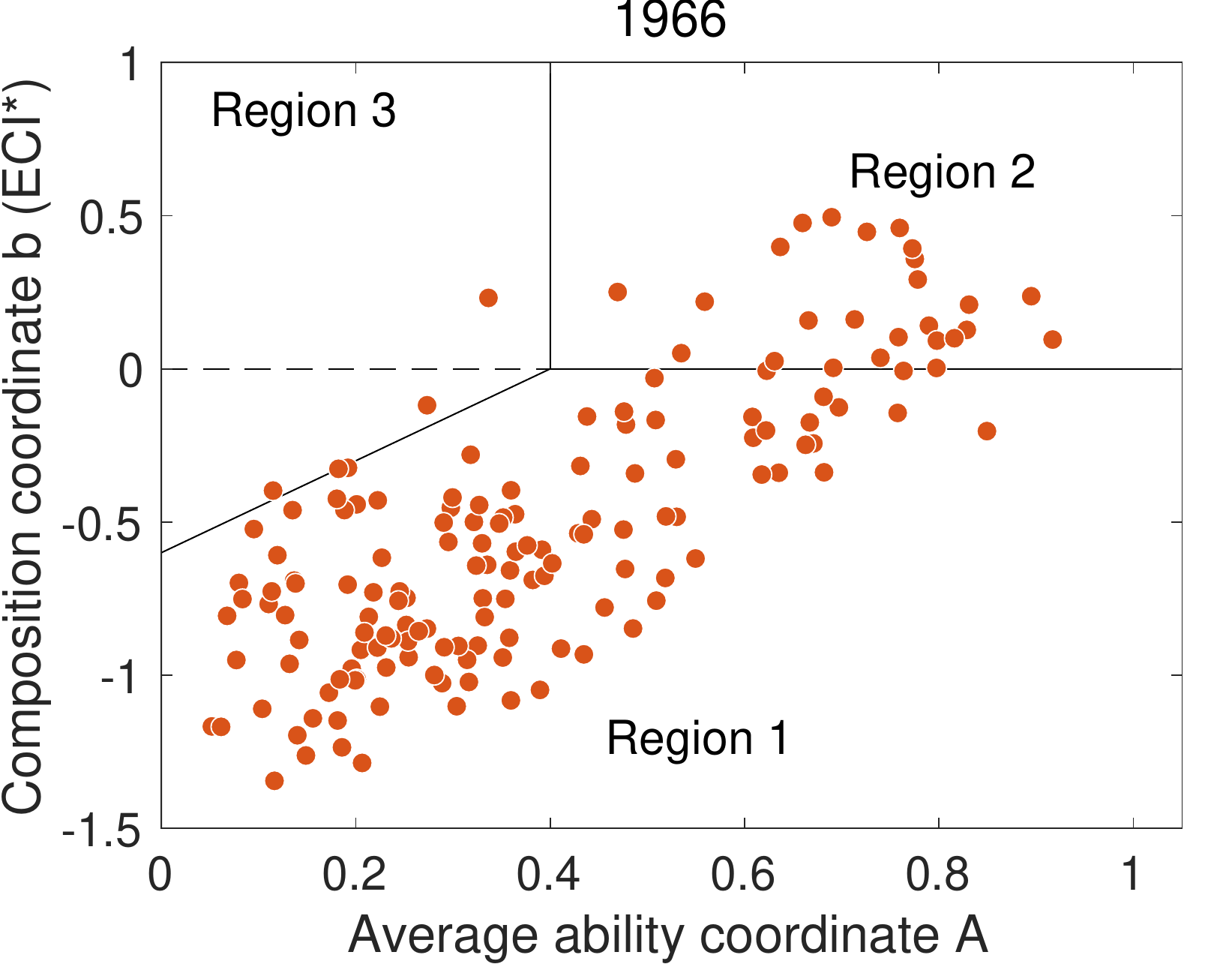}\hspace{0.25in}
\includegraphics[width=0.35\textwidth]{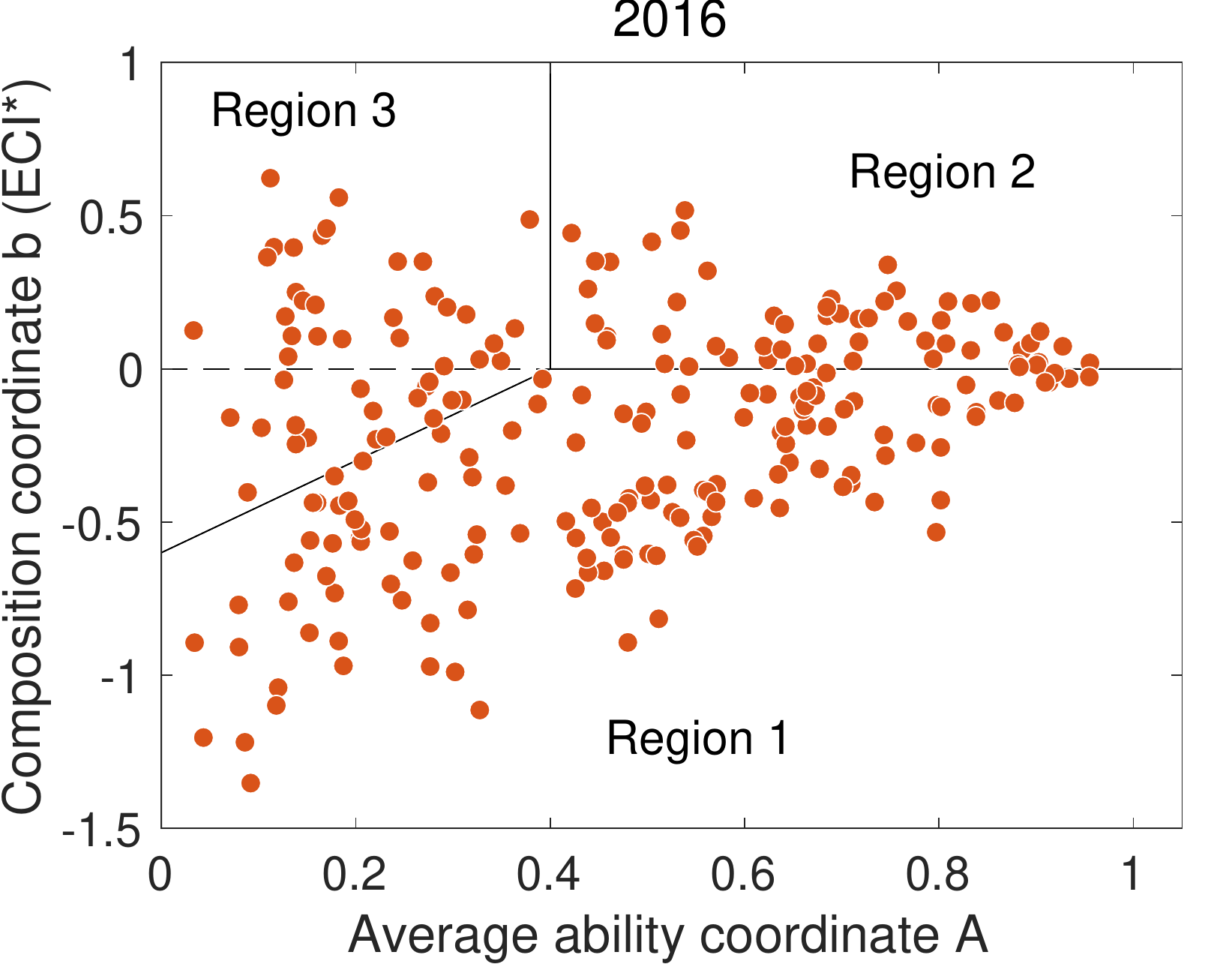}
}
\caption{\ed{\textbf{a} Correlations over time among our $A$ and $b$ (ECI*) coordinates, complexity metrics, and related quantities. \textbf{b} Scatter of regions in the $A$-$b$ plane in 1966 and 2016.}}
\label{fig_correlationsOverTime}
\end{figure}

\ed{Correlations within the two groups of diversity-like and composition-like metrics were high across time (Fig. \ref{fig_correlationsOverTime}a).  In addition, we note that  correlations across these two groups are high in early years of our data, where regions overwhelmingly lie in a diagonal band (Fig. \ref{fig_correlationsOverTime}b) such that a region’s $A$ coordinate is strongly related to its $b$ coordinate.  We are mainly interested in correlations within each group of metrics, and do not determine the extent to which these changes in cross-group correlations derive from aspects of our data (e.g. new countries becoming available in our dataset over time) or from changes in the nature of development.  An exception to the high within-group correlations is $X_2 / \sqrt{d}$, which in early years is weakly correlated with other compositional quantities. GENEPY was originally studied on data in the years 1995 - 2017, and thus we are examining this quantity in a period well outside the one in which it was tested.}

\FloatBarrier

\FloatBarrier
\section{Supplementary note: Dimensionality reduction}
We can relate the theory here to recent works that study the geography of economic activity and complexity metrics through the lens of dimensionality reduction.  For instance, Mealy et al. (2019) \cite{Mealy2019} point out that the conventional ECI can be seen as a dimensionality reduction tool, equivalent to a spectral clustering algorithm that orders countries along an axis in a way that tries to preserve the proximities between baskets of activities.  Similarly, Sciarra et al. (2020) \cite{Sciarra2020} put forward a complexity index that reduces the dimensionality of economic activity data by combining information from two eigenvectors of a country-country similarity graph.  Brummitt et al. (2020) \cite{Brummitt2020}, in turn, apply machine learning methods to extract dimensions that characterize variation in historical export baskets across countries and time, finding axes that strongly correlate with the PCIs.  In the works above, dimensionality reduction involves strategies to summarize complex data.  In contrast, in exploring our dynamical model, we used the tool of eigenmode decomposition. Eigenmode decomposition is not focused on discovering structure in data per se, but on describing a dynamical process.  Nevertheless, it is well understood that these approaches are related, and here we briefly summarize the relationship.

First, the outcome of eigenmode decomposition is very similar to dimensionality reduction techniques: It returns dimensions (the eigenvectors $\vec{v}_\mu$) in which one can describe high-dimensional data (the matrix of activity baskets $\vec{R}(t)$), along with the coordinate projections of each data point onto these axes (the coefficients $c_\mu(t)$). The two approaches differ in how directions in the vector space are determined and interpreted.  Typically, dimensionality reduction techniques aim to identify directions that capture large amounts of variation in data, while eigenmodes characterize coherent dynamical patterns.  This difference is closely related to the distinction between proper orthogonal decomposition (POD) and dynamical mode decomposition (DMD) in applications of \ed{machine} learning to dynamical systems (e.g. \cite{Tu2014}): Whereas POD modes optimally reconstruct a data set of snapshots of a system over time, DMD modes identify coherent structures in the dynamics.

\begin{figure*}[t]
\center
\includegraphics[height=0.24\textheight]{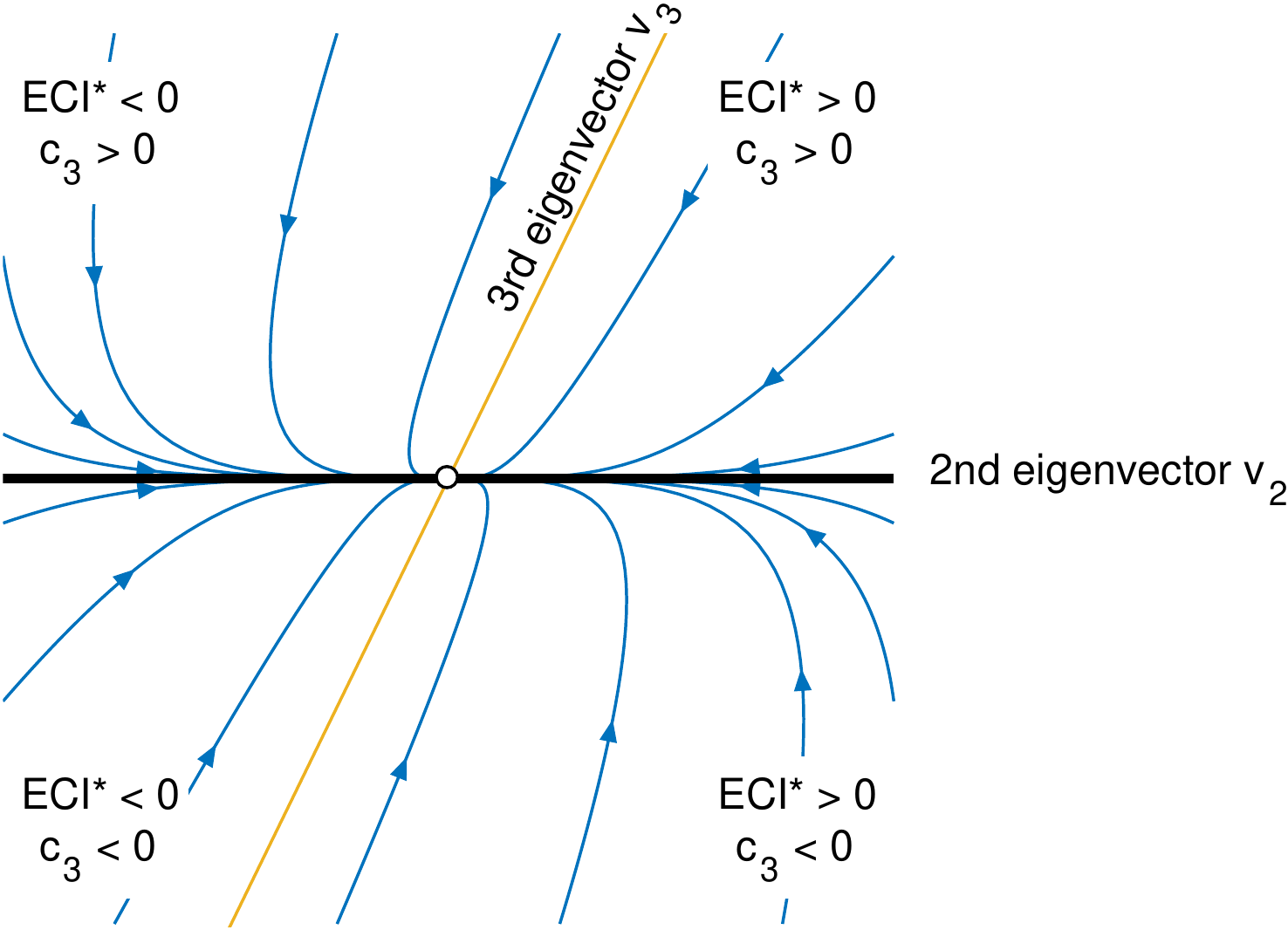}
\caption{An illustration of \ed{the} model dynamics.  We display the phase portrait in the 2D plane corresponding to the 2nd and 3rd eigenmodes in a case where $\tau_2 / \tau_3 = 4$.  Different points in the plane correspond to different coordinate projections for an activity basket $\vec{R}(t)$ onto the 2nd and 3rd eigenvectors $\vec{v}_2$ and $\vec{v}_3$.  We present a skewed coordinate system as a reminder that these vectors in general are non-orthogonal, $\vec{v}_2 \cdot \vec{v}_3 \neq 0$.  The four quadrants are labelled by whether the projection onto each eigenvector is positive or negative.  Blue lines show the paths of a country that obeys the model depending on its starting position.  The model has a stable fixed point where the projections onto both eigenvectors are zero.  Because the speed of convergence is much faster along the 3rd eigenvector, however, variation along this direction quickly shrinks, while variation along the 2nd eigenvector is sustained for a longer time.}
\label{fig_ECI_as_coordinate}
\end{figure*}

The eigenmode decomposition not only generates a representation of data in a vector space, but shows that the model predicts that countries' activity baskets will converge to a lower-dimensional subspace that corresponds to the modes with the largest time scales.  This is a well-known outcome of the interaction of structure and dynamics in networks \cite{Simon1961,Schaub2019}.  Heterogeneity in the time scales of different eigenmodes leads countries' activity baskets to collapse onto a low-dimensional subspace of possible activity baskets as differences between countries shrink first among quickly-decaying dimensions, leaving activity baskets scattered primarily along slowly-decaying dimensions.  A simple example is given in Fig. \ref{fig_ECI_as_coordinate}, which depicts the phase portrait of the model if we assume that the characteristic time scale of the 2nd eigenmode is four times that of the 3rd, $\tau_2 / \tau_3 = 4$.  Phase lines show the path of a country that obeys the model, and show that variation along the 3rd eigenvector shrinks much more quickly than variation along the 2nd eigenvector.  This illustrates how the theory here can be viewed through a dimensionality reduction lens, and indeed encourages such a perspective.

\FloatBarrier
\section{Supplementary note: Benefits of seeing complexity metrics as coordinates of \mbox{structural} change} \label{sec_benefits}
In discussing our dynamical modeling framework, we emphasized that quantities very similar to complexity metrics (the $A$ and $b$ coordinates) can be motivated using standard arguments of dynamical systems theory.  These arguments lead to measures that have little to do with `complexity' per se, calling into question the usual interpretation of complexity metrics.  However, we could also ask what benefits or opportunities are suggested by the framework here, which invites us to see complexity metrics as quantities that track the path of structural change by summarizing an economy's basket of activities in interpretable, low-order terms.  We describe four possible benefits below.

First, a theory like the one here can \ed{potentially} give interpretations to the signs and magnitudes of some complexity metrics.  For example, the conventional PCI and ECI are usually standardized to have zero mean and unit variance, implying that only the way they rank products and countries matters.  Yet our theory assigns clear meaning to signs and magnitudes: Elements of $\vec{v}_2$ ($\vec{PCI}^*$) convey which products are vacated or reached, and how large the change in these products is as structural change occurs.  Similarly, $b(t)$ (ECI*) conveys more than just rank information -- its magnitude conveys how far a country has progressed in a particular direction of change.\footnote{\ed{As a caveat, we note that there are still numerical and sign differences among these metrics and our coordinates, due to differences in how each one is constructed, including choices about the product proximity matrix selected; the use of RCAs as measures of ability; and how RCAs are transformed.}}

Second, interpreting complexity metrics as coordinates raises the possibility of tracking countries' structural change in absolute terms.  A longstanding concern of research on economic growth is whether countries are converging in income.  A related question is whether countries are converging in industrial composition.  It is clearly not possible to track such a convergence if complexity metrics can only provide rank information about countries.  But it is possible to do this using the $A$ and $b$ coordinates, or any complexity metric, if one can justify that the cardinal values of these metrics are meaningful, so that different countries can potentially reach similar coordinate values over time.

Third, by deriving these coordinate measures from a dynamical model of the process of development, improving the description of this process -- e.g. developing better proximity matrices $\Phi$, or improving on the dynamical model presented in the main text -- can feed into better summary measures of structural change, offering a path for refinement of these measures.

Finally, because our coordinates depend on the network of products $\Phi$, our results raise the question of what happens as this network changes.  \ed{While it is conventional to assume that $\Phi$ is fixed for the purposes of predicting product transitions, p}roximities between different products can evolve \ed{\cite{hidalgo2009dynamics}} as the technologies underpinning these products change, which in turn could shift the directional tendency of structural change.  \ed{While in the past countries have undergone well-documented patterns of development (e.g. the shift out of agricultural products toward manufactured goods \cite{Imbs2003,Brummitt2020}) this canonical trajectory could change.  A framework like the one here can account for such shifts in a similar way that index numbers do in other settings.  Change in a region's $b(t)$ coordinate (for example) would be split into two components -- one due to changes in a country's weights on the elements of $\vec{v}_2$, and one due to changes in the elements of $\vec{v}_2$ themselves.  This would allow one to decompose movements of a country into sum of a long-term movement along the direction of development, and shifts in the direction of development itself,} whereby the relative sequence of different products can rise and fall over time.  \ed{This would also be one way for studies to evaluate the quality of the approximation that $\Phi$ is fixed in time, as this would be reflected in the rate of change of $\vec{v}_2$.}

\section{Supplementary note: On the ECI* weights when $\Phi = \Phi^{P}$}
\ed{As discussed in the Methods, the special case $\Phi = \Phi^P$ is noteworthy} because $\vec{PCI}^*$ coincides with $\vec{PCI}$.  \ed{There are also reasons why this special case is interesting for the ECI*.}  \ed{First, note that in this case $L_\Phi$'s first left eigenvector $\vec{w}_1 = \vec{\pi}$ is proportional to the vector of product ubiquities:}
\begin{align}
\vec{\pi} \propto \vec{u}.
\end{align}
To see this, observe that $L_\Phi = I - \tilde{\Phi}$ and $\tilde{\Phi}$ have the same eigenvectors, and that $\vec{u}$ is a left eigenvector of $\tilde{\Phi}^P = U^{-1} M^T D^{-1} M$ with eigenvalue 1:
\begin{align}
\vec{u}^T \P 
= \vec{u}^T U^{-1} (M^T D^{-1} M) 
= \vec{1}^T M^T D^{-1} M 
= \vec{d}^T D^{-1} M 
= \vec{1}^T M 
= \vec{u}^T.
\label{eq_firstLeftEV_ofScriptP}
\end{align}
Taking $\vec{\pi} \propto \vec{u}$, the ECI* (Eq. (9) of the main text) becomes
\begin{align}
\text{ECI}_c^*(t) = \sum_p \left( \frac{R_{cp}(t) u_p}{\sum_{p'} R_{cp'}(t) u_{p'}} \right) \text{PCI}_p.
\label{eq_ECI_forPhiP}
\end{align}
\ed{Thus,} in this special case, the weights used by the ECI* are ubiquity-adjusted RCAs.  These weights emerge as a by-product of using the proximity matrix $\Phi = \Phi^P$, and do not represent an intentional choice about how different products in the export basket should be weighted.  However, such a ubiquity adjustment could nevertheless be beneficial from a measurement standpoint, because it puts more weight on products for which high RCAs are harder to achieve and are therefore more exceptional.  For example, many countries in the world produce textiles, but only a few produce ostrich eggs.  As a result, almost all ostrich egg-producers have high RCAs in this product.  In contrast, to achieve a high RCA in a high-ubiquity product such as textiles, a country must outcompete many others.  In general, achieving high RCAs will be more difficult in ubiquitous products than in nonubiquitous ones, and it would be reasonable to regard high values of $R_{cp}$ as more impressive the more ubiquitous product $p$ is. The conventional ECI not only thresholds RCAs in weighting products in a country's activity basket, but it also makes no adjustment for how competitive each product's export market is.  In contrast, the ECI* uses unthresholded RCAs, and in the special case $\Phi = \Phi^P$, it emphasizes the RCAs of ubiquitous products.  We do not offer a more general interpretation of the ergodic weights $\pi_p$ that applies outside the special case $\Phi = \Phi^P$, but exploring such issues could reveal interesting aspects of measurement in those contexts as well.

\clearpage